\documentclass[reprint,twocolumn, superscriptaddress,
 amsmath,amssymb,
 aps,
 longbibliography,
 footinbib,
]{revtex4-2}

\usepackage{graphicx}
\usepackage{dcolumn}
\usepackage{bm}
\usepackage[dvipsnames]{xcolor}
\usepackage{comment}
\usepackage{float}
\usepackage[colorlinks=true,bookmarks=false,citecolor=blue,urlcolor=blue]{hyperref}

\let\vec=\mathbf

\begin{document}

\title{Multipolar theory of bianisotropic response}

\author{Maria Poleva}
\author{Kristina Frizyuk}
\author{Kseniia Baryshnikova}
\author{Andrey Bogdanov}
\email{a.bogdanov@metalab.ifmo.ru}
\author{Mihail Petrov}
\email{m.petrov@metalab.ifmo.ru}
\author{Andrey Evlyukhin}

\begin{abstract}
    Bianisotropy of metaatoms is usually associated with their nonlocal response and the mutual coupling between electric and magnetic dipole moments induced by the incident field. In this work, we generalize the theory of bianisotropy beyond the dipole response to the cases of arbitrary high-order multipole resonances. We demonstrate that bianisotropy is exclusively  caused by the  absence of the inversion symmetry of metaatoms. The strength of the  bianisotropy response is normally increased with the size of a meta-atom but its origin is fully connected to the symmetry of the structure. As an important example of bianisotropic particle, we consider a triangular prism and show how accounting for the higher-order multipoles prevents the violation of the Onsager-Casimir conditions for kinetic coefficients appearing in the dipole approximation. The developed theory is an important step towards a deeper insight into the scattering properties of nanoantennas and metaatoms.   
\end{abstract}

\maketitle

Metasurfaces, metamaterials, and nanoantennas are the main building blocks of nanophotonics allowing giant field enhancement and precise control over light at the subwavelength scale. Their optical properties are strongly governed by the  response of {\it meta-atoms} -- their elementary units. The accurate description of light interaction with individual meta-atoms and their small clusters is of high interest to the nanophotonics community \cite{jung2016electromagnetically, terekhov2017multipolar, koshelev2020dielectric} as it opens a way for effective engineering the nanophotonic structures with predesigned optical properties. One of the most powerful tool for such an analysis is the {\it multipolar expansion} method. Within this method \cite{Alaee2019Jan}, the electromagnetic field inside and outside the scatterer of an arbitrary shape is described by a superposition of vector spherical harmonics (VSHs) forming a complete set of basis functions known as {\it multipoles}~\cite{waterman1961multiple}. In the scattered field domain, the meta-atoms can be considered as  effective point scatterers with  specific polarizabilities corresponding to its dominant multipole moments. As soon as the meta-atoms are usually sufficiently smaller than the wavelength, the scattered field can be well described by only the  electric  (ED) and magnetic  (MD) dipoles that makes it extremely helpful to interpret and predict the optical properties of single meta-atoms and their ensembles \cite{evlyukhin2010optical}. On this way, many fascinating effects have already been demonstrated and explained such as non-radiating anapole effect \cite{miroshnichenko2015nonradiating,zenin2017direct,baryshnikova2019optical,yang2019nonradiating,savinov2019optical},  directional scattering \cite{liu2018generalized,shamkhi2019transverse}, cloaking and invisibility effects \cite{fleury2015invisibility,babicheva2021multipole}.

The electric and magnetic dipole moments ($\mathbf{p}$ and $\mathbf{m}$) of meta-atoms are not always independent. They can be mutually coupled, i.e. the both components of incident electromagnetic field ($\mathbf{E}$ and $\mathbf{H}$) simultaneously induce $\mathbf{p}$ and $\mathbf{m}$. This effect is usually called {\it bianisotropy} (BA) and can be formally described as follows~\cite{Alaee2015-All-dielectricrecipr,Bobylev2020-Nonlocalresponseof,evlyukhin2020bianisotropy,Albooyeh2016Dec}:
\begin{eqnarray}\label{eq1}
\begin{split}
{\bf p}& =&{\varepsilon_0}\hat{\alpha}^{\rm E E}{\bf  E}+\frac{1}{c}\hat{\alpha}^{\rm E H} {\bf H}\label{eq1a},\\
{\bf m}& =&\hat{\alpha}^{\rm H H}{\bf H}+\frac{1}{Z_0}\hat{\alpha}^{\rm H E} {\bf E}\:,
\end{split}\label{eq1b}
\end{eqnarray}
where $\bf E$ and $\bf H$ are the electric and magnetic fields of the incident wave at the dipoles coordinates,  $c$ is the speed of light,    $Z_0$ is the wave impedance, $\varepsilon_0$ is the vacuum permittivity, and  $\hat{\alpha}^{\rm EE}, \hat{\alpha}^{\rm EH}, \hat{\alpha}^{\rm HE}, 
\hat{\alpha}^{\rm HH}$ are the second rank polarizability tensors. 

\begin{figure}[t]
\centering
\includegraphics[width=1\columnwidth]{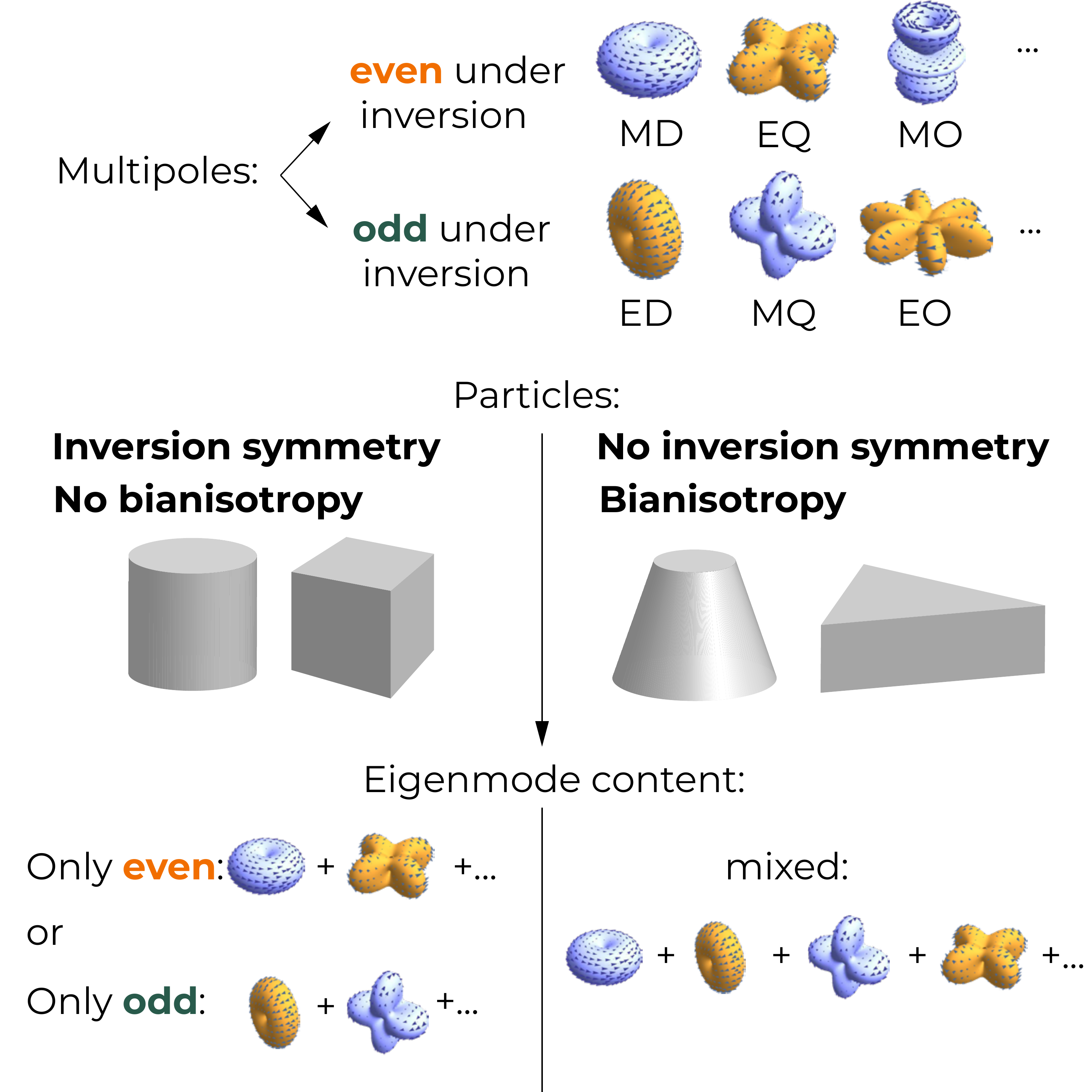}
\caption{General idea of multipolar bianisotropy. Vector spherical harmonics can be divided into even and odd under inversion operation. The examples of even and odd multipoles are given in the head of the figure. If a particle is symmetric under the inversion (left column), then it has no bianisotropic response and  its eigenmodes consist of either only odd or even multipoles. As soon as the particle does not possess inversion symmetry (right column), it has bianisotropic response and its eigenmodes consist of the multipoles of both parities.
} 
\label{fig1}
 \end{figure}
One needs to stress that the BA contribution is negligible in metaatoms much smaller than the wavelength, while it becomes more apparent in the structures with sizes comparable to the wavelength. And that has  already led to appearance of many fascinating effects in nanophotonics such as polarization control with optical metasurfaces, realizing  topological photonic states, broken reciprocity, unusual optical forces,  and others \cite{asadchy2018bianisotropic,Rodriguez2014Dec,Achouri2020Aug,Wei2022Mar}. At the same time, the higher order resonances start playing the key role in the metaatoms with sizes comparable to the wavelength~\cite{Wu2020Jan,Muhlig2011Jun,Gladyshev2020Aug,Alaee2017Jan}  allowing to generalize many optical  effects such as Kerker effect~\cite{shamkhi2019transverse,liu2018generalized}, anapole effects~\cite{grinblat2017efficient,zenin2017direct}, increased Q-factors of resonant modes~\cite{bogdanov2019bound,odit2021observation}.  The selective excitation of the solely higher order resonances without coupling to the lower ones is possible by using specific configurations of the incident optical of beam~\cite{sakai2015excitation,wozniak2015selective,Das2015Dec}. Moreover, for some specific shapes of scatterers, the dipole model [Eqs.~\eqref{eq1b}] does not describe the electromagnetic response correctly and even results in the violation of the Onsager-Casimir conditions for kinetic coefficients~\cite{Proust2016Sep,Poleva2021Nov}. In this sense, the generalization of the BA effects for higher-order resonances is a natural step in the development of the multipole theory and all-dielectric metaphotonics.



 In this Letter, we reconsider and generalize the concept of  bianisotropy to meta-atoms of an arbitrary shape and size supporting  high-order multipole resonances. We show that the origin of bianisotropy is fully defined by the symmetry of the meta-atom rather than its size. The size of the meta-atom, i.e. nonlocal response (spatial dispersion), is only responsible for the strength of BA but not for its presence. Applying the group theory formalism, we identify the selection rules enabling the multipolar bianisotropy effects and enhanced with resonant response. With this knowledge, the predefined engineering of the bianisotropic response of arbitrary shaped scatterers becomes possible and pays the way towards great prospective in nanophotonics. The developed approach is consistent with all previous models of BA but brings a more solid and generalized view to the  well-known electromagnetic BA effect.    
 

\begin{figure}[t]
\centering
\includegraphics[width=1\columnwidth]{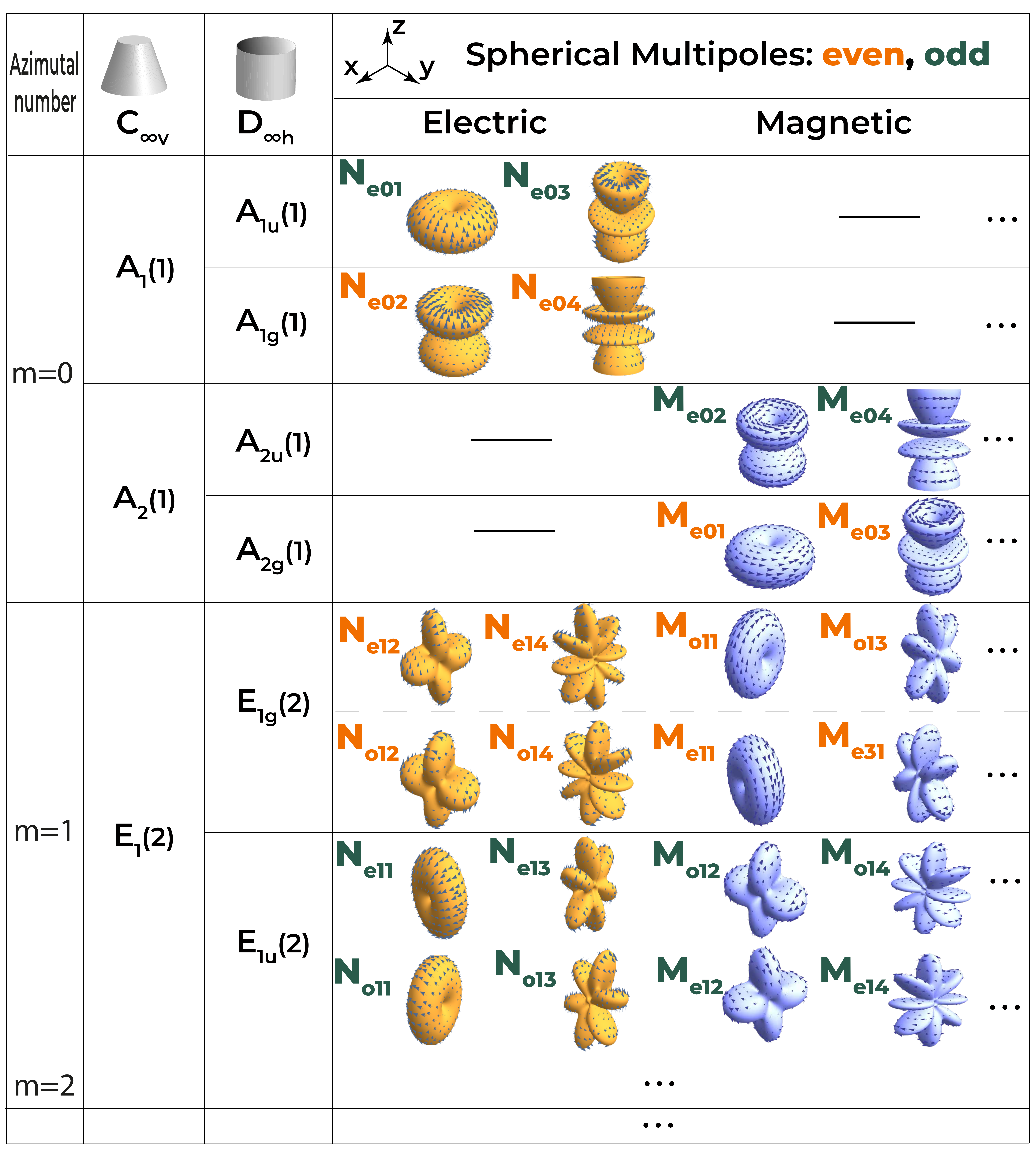}
\caption{Multipolar decomposition of eigenmodes of an object with the symmetry groups $C_{\infty v}$ and $D_{\infty h}$. In the left-hand column, there are azimuthal numbers m.{In the column below the symmetry group identifier, there are the irreducible representations of the group and their dimensions in the brackets.} In the right-hand column, there are first-orders multipoles (VSHs) of the eigenmodes associated with the corresponding irreducible representation. The electric VSHs are depicted in orange, the magnetic VSHs are in blue. The letters corresponding to multipoles even under inversion are orange, names of odd multipoles are green. Figure shapes represent the distribution in the far-field zone the absolute value of the corresponding multipole electric field, the arrows depict the polarization of the field emitted in the given direction.}
\label{fig:cone_cyl}
 \end{figure}

Historically, BA was introduced for moving media~\cite{cheng1968covariant,cheng1968time}. In such systems, electric and magnetic displacements $\mathbf{D}$ and $\mathbf{H}$ are the functions of both electric and magnetic fields $\mathbf{E}$ and $\mathbf{B}$. One can show that in the general case such a connection is possible only in the media without inversion symmetry~\cite{Fan2013Apr,Bobylev2020-Nonlocalresponseof}. Therefore, a BA particle being a structural element of BA medium can be defined as one that has {\it no inversion symmetry}. Some of the well-known examples of BA particles are shown in Fig.~\ref{fig1}. Absence of the inversion symmetry  immediately affects the multipolar content of the field scattered by the particle. Indeed, the multipoles entering an eigenmode are transformed as well as the eigenmode under all transformations of particle's symmetry group. Thus, for the particles with inversion symmetry, all the eigenmodes can be classified into to even and odd with respect to the inversion operation. Thus, each eigenmode consists of either all even or odd multipoles. For BA particles, we may conclude that {\it each eigenmode contains multipoles of both parities}, which  is illustrated in Fig.~\ref{fig1}.                

The specific multipolar content of eigenmodes can be found using the Wigner theorem that states that all the modes in a resonator are transformed by irreducible representations of the resonator's symmetry group~\cite{ivchenko2012superlattices,Gladyshev2020Aug,Xiong2020Feb,Tsimokha2021-Acousticresonators,Piroth2007-FundamentalsoftheP}. The mode consists of the multipoles, which transform by the same irreducible representation as the mode itself. We will follow the notations of Ref.~\onlinecite{Bohren1998Mar} describing the multipoles by the VSHs $\vec{M}_\beta$ (magnetic harmonic) and $\vec{N}_\beta$ (electric harmonic). Index  $\beta=\{e/o,m,\ell\}$ encodes the parity of the harmonic (even '$e$' or odd '$o$'), the total angular momentum $\ell=1,2,3,...$ and its projection $m=0,1,...,\ell$. We should note that indices  '$e$' and '$o$' shows the parity with respect to $\phi\rightarrow-\phi$ but not to the inversion \cite{supp_mat}.


The table in Fig.~\ref{fig:cone_cyl} shows the eigenmode classification and their multipole content for a truncated cone (C${_{\infty v}}$ symmetry group) and a cylinder (D${_{\infty h}}$ symmetry group) which are known as a two distinct examples of BA and non-BA particles. The multipolar content for other shapes can be found in Ref.~\onlinecite{Gladyshev2020Aug}. One can see that for the cylindrical particles, the modes are classified by azimuthal number $m$ and parity with respect to the inversion. For $m=0$, all the modes are additionally  subclassified into electric and magnetic ones. We use additional  orange/green coloring to distinguish between the even/odd multipoles. The modes with $m\geq1$ are doubly degenerate due to cylindrical symmetry and their linear combinations can be associated with the clockwise and counterclockwise rotation directions. The transformation of the cylinder into the truncated cone mixes the odd and even modes of the same azimuthal number $m$. Therefore, the parity is no longer the quantum number. Indeed, let us consider the illumination of a cylinder  by a plane wave propagating along the $z$-direction and polarized along the $y$-axis (see multipolar decomposition of planes wave with different polarizations in Fig.~S1 of Ref.~\cite{supp_mat}). In this case, the wave will excite $\mathbf{N}_{o11}$ harmonic from E$_{1u}$ and  $\mathbf{M}_{e11}$ harmonic from E$_{1g}$. They have different parity and are excited independently -- the magnetic field of the plane wave excites only the modes of E$_{1g}$ and electric field excites only the modes of E$_{1u}$. However, when the cylinder is transformed  to a truncated cone, both $\mathbf{N}_{o11}$ and  $\mathbf{M}_{e11}$ enter the same representation, and their  cross excitation according to  Eqs.~\eqref{eq1b} becomes possible leading to the BA response. The strength of BA can be characterised  by the corresponding T-matrix element~\cite{supp_mat}.

\begin{figure}
 \includegraphics[width=1\linewidth]{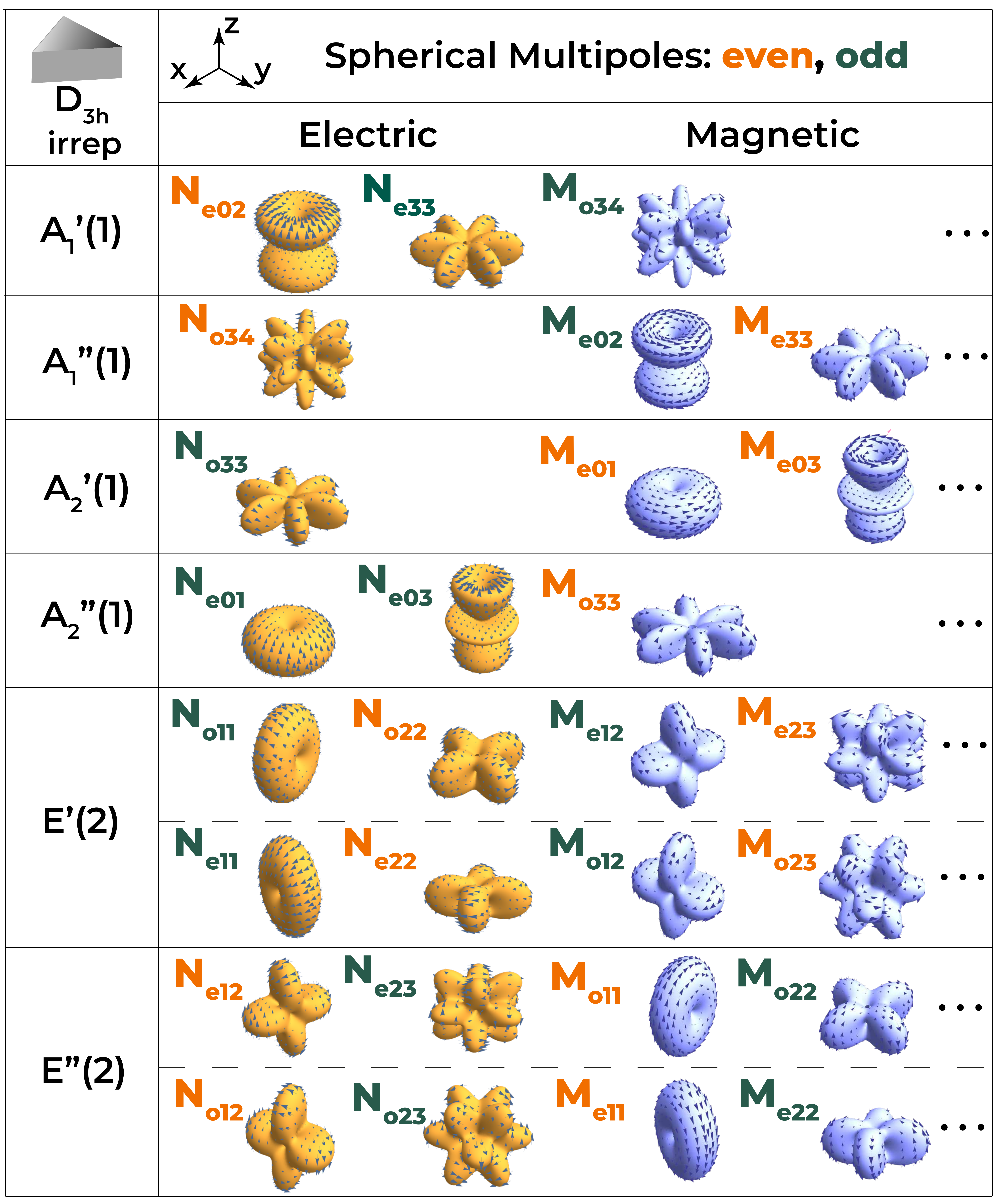}
    \caption{Multipolar decomposition of eigenmodes  of an object with the symmetry group  $D_{3h}$, corresponding to the prism particles. {In the left-hand column of the table there are the irreducible representations of the group and their dimensions in the brackets.} In the right-hand column, there are first-order electric (yellow) and magnetic (blue) multipoles (VSHs) of the eigenmodes associated with the corresponding irreducible representation. The figures shows the field amplitude in the far-field zone, while the arrows show the polarization of the field. Orange and green coloring of letters corresponds to  even and odd inversion parity.  }
    \label{fig:prism}%
\end{figure}

Understanding the multipole content of resonators allows immediate extension the BA to the domain of the {\it higher-order multipoles}, that become essential for larger sizes of the resonator when the role of nonlocality (spatial dispersion) increases. Indeed, one can see from Fig.~\ref{fig:cone_cyl}(a) that the electric and magnetic quadrupole multipoles also  enter the same representation E$_1$  of cone modes, thus, enabling dipole-quadrupole and  quadrupole-quadrupole BA, which has been observed in various photonic and plasmonic systems~\cite{BernalArango2014May,Mun2019May,Proust2016Sep}. One should also note that the common understanding that the BA is a coupling between the electric and magnetic resonances is not true in general. Indeed, as it can be seen from Fig.~\ref{fig:cone_cyl}, the eigenmodes A$_1$ and A$_2$ of a truncated cone contain the opposite parity multipoles with $m=0$ of only one type either magnetic or electric. These modes can be excited independently, for example, by azimuthally or radially polarized Bessel beam~\cite{Das2015-Beamengineeringfor,Melik-Gaykazyan2018-SelectiveThird-Harmo,Koshelev2020Jan}. Such type of BA was not consider previously and remains the subject of further research. 

In this prospective, a non-trivial example of  isosceles triangular prism is of special interest as the standard dipole model described by Eqs.~\eqref{eq1b}] is not applicable resulting in  the violation of the Onsager-Casimir conditions for the kinetic coefficients [see Fig.~\ref{fig:prism}]~\cite{Poleva2021Nov}. As a non-centrally symmetric particle it should formally possess BA response, however one can see that electric and magnetic dipole components never enter the same representation: the $x$- and $y$-oriented EDs ($\mathbf{N}_{e11}$ and $\mathbf{N}_{o11}$) enter E$^{\prime}$, while $x$- and $y$-oriented MDs ($\mathbf{M}_{e11}$ and $\mathbf{M}_{o11}$)  enter E$^{\prime \prime}$. The same happens for the $z$-oriented dipoles \cite{supp_mat}. Thus, formally the dipole-dipole BA is forbidden, while dipole-quadrupole and higher order BA is allowed.      

  \begin{figure}
 \includegraphics[width=1\linewidth, height=1\linewidth]{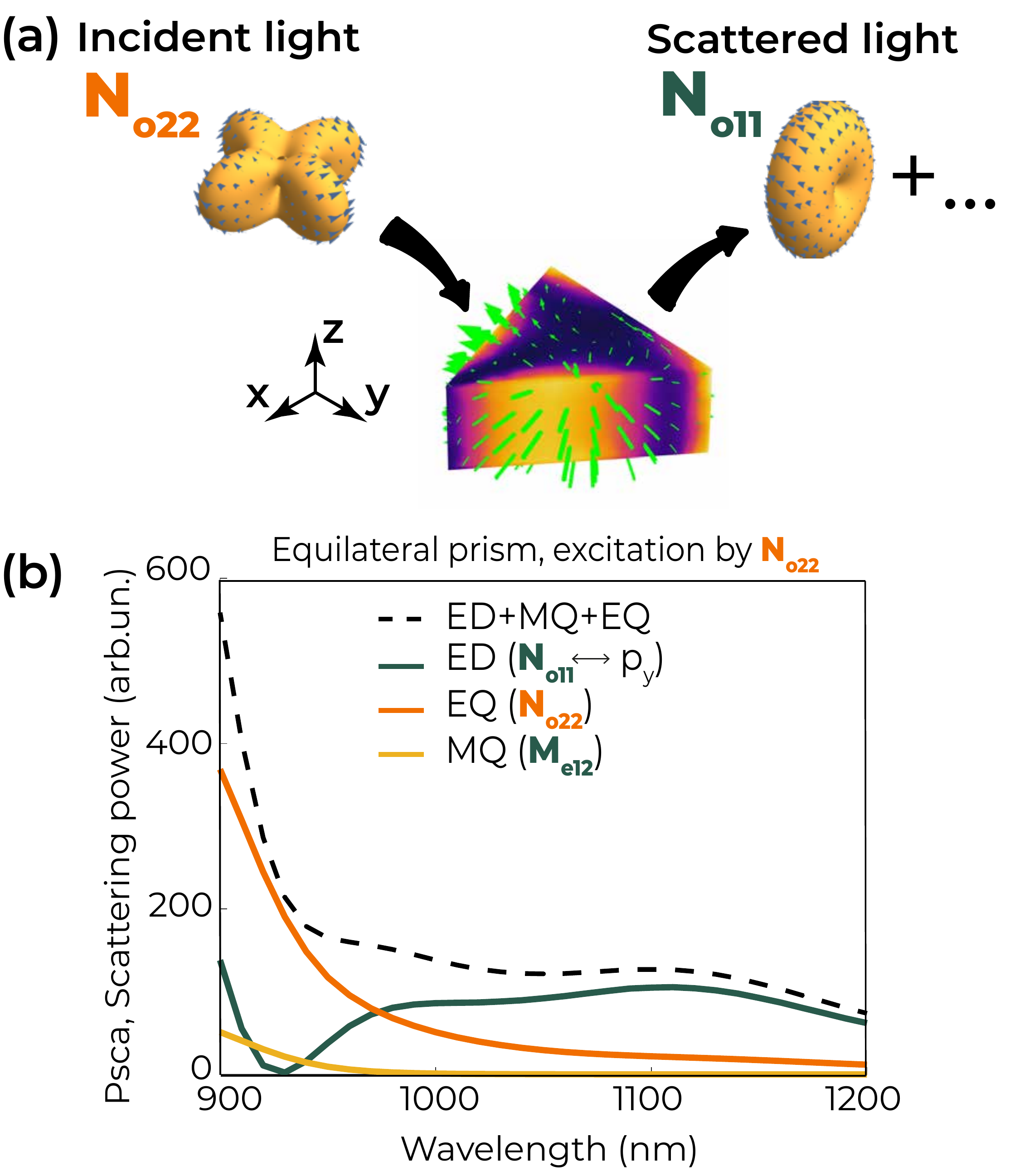}
    \caption{(a) Schematic depiction of the dipole-quadrupole BA. Green arrows on the prism denote the direction and amplitude of the electric field excited by the field, which multipole decomposition consist of an electric quadrupole $\vec N_{o22}$. The color represents the distribution of the module of the electric field. (b) The spectrum of the total and partial scattering powers at the wavelength $\lambda=1050$~nm under the illumination of an electric quadrupole $\vec N_{o22}$. }
    \label{coupling}%
\end{figure}
Indeed, the electric quadrupole  $\mathbf{N}_{o22}$ and the electric dipole $\vec N_{o11}$ are both present at the same mode corresponding to E$^{\prime}$ representation [see Fig. \ref{fig:prism}]. Hence, if the incident field contains $\vec{N}_{o22}$, then the modes from E$^{\prime}$ are excited and the scattered field will contain the electric dipole $\vec{N}_{o11}$ which has the opposite parity. This manifestation of higher-order BA is shown schematically in Fig. \ref{coupling}(a). To demonstrate this effect numerically, we consider an isosceles triangular prism in air made of a material with dielectric permittivity $\varepsilon_p=12.67$ that is close to the silicon in the near infra-red spectral range \cite{palik1998handbook}. The geometric parameters are listed in the caption. The prism is excited by the spherical wave whose electric field contains only the harmonic $\vec N_{o22}$.    Figure~\ref{coupling}(b) shows the spectra of the total and partial scattering powers corresponding to $\vec {N}_{o11}$, $\vec {N}_{o22}$, and $\vec {M}_{e12}$ calculated in COMSOL Multiphysics\texttrademark. One can see that the prism scatters the incident field proportional to even harmonic $\vec {N}_{o22}$ into the channels corresponding to odd harmonics   $\vec {N}_{o11}$  and $\vec {M}_{e12}$ manifesting the higher-order BA.

The standard BA model [Eqs.~\eqref{eq1b}] can be directly extended beyond the dipole terms via including the spatial derivatives of the electric and magnetic field that is equivalent to the accounting for the higher-order multipoles in the incident field~\cite{Varault2013-Multipolareffectson,Achouri2021Feb}
\begin{equation}\label{pi,mi}
\begin{gathered}
{p_i} ={\varepsilon_0}{\alpha}^{\rm E E}_{i j}{E_j}+\frac{1}{c} {\alpha}^{\rm E H}_{i j} {H_j}+
\frac{\varepsilon_0}{2k_0}
{A}^{\rm EE}_{ijk}(\nabla_k {E}_{j})+\dots,\\
{m_i} ={\alpha}^{\rm H H}_{i j}{H_j}+\frac{1}{Z_0}{\alpha}^{\rm  H E}_{i j} {E_j}+
\frac{1}{2 Z_0 k_0}
{A}^{\rm H H}_{ijk}(\nabla_k {H}_{j})+\dots.
\end{gathered}
\end{equation}
Here  $\hat A^{\rm EE}$ and $\hat A^{\rm HH}$ {are} the non-local polarizability tensors of third rank, $k_0$ is the vacuum wavenumber. Applying the model provided Eq.~\eqref{pi,mi} to the case of scattering of a plane wave propagating along $y$-axis with $x$-polarization, for which  $\alpha_{yx}^{\rm EE}=0$ and  $\alpha_{yz}^{\rm EH}=0$, one can write 
\begin{equation}\label{eq:General_Bianis}
    p_y=a_{yxy}\left.\left[\frac{\partial}{\partial y}(E_x e^{ik_y y})\right]\right|_{r=0}+\dots=a_{yxy}ik_yE_x+\dots,
\end{equation}
where $a_{yxy}=\varepsilon_0 A_{yxy}^{\rm EE}/2k_0$. The field component on the right-hand  side of Eq.~\eqref{eq:General_Bianis} corresponds to the electrical quadrupole  $\vec{N}_{o22}$ contained both in the incident and scattered fields. However, one should note that this model [Eqs.~\eqref{pi,mi}] is not complete within the quadrupole approximation as it accounts for the inducing the ED by EQ but it does not accounts for the reciprocal effect. A more natural formalism for accounting for the cross-coupling between the different multipoles is the use of T-matrix which coefficients $T_{\beta_{\text{out}},\beta_{\text{in}}}^{W_{\text{out}},W_{\text{in}}}$ show the connections between the complex amplitudes of the incident and scattered fields. Here, the the index $W_{\text{in}/\text{out}}=\{N,M\}$ encodes the type of harmonic (electric or magnetic) and  index $\beta_{\text{in}/\text{out}}=\{o/e,m,n\}$. For the case considered in Fig.~\ref{coupling}(a), one can find the following relation (see Eq.~(S44) in Ref.~\cite{supp_mat}):      
\begin{equation}\label{a_N}
a_{yxy}=-\frac{5 i \pi}{2 \sqrt{3}  k_0 \omega} T^{NN}_{o11,o22},
\end{equation}  
that is valid for small particles. The general relation between T-matrix elements and multipolar polarizability coefficients can be found in Ref.~\cite{Mun2020May}.

Finally, we need to stress that the group analysis allows finding the selection rules for the cross-coupling between the multipoles but it does not predict the strength of the BA response. The strength of the BA is described by the off-diagonal elements of the corresponding T-matrix elements. They can be found, for instance, using the resonant state expansion method~\cite{Doost2014Jul, supp_mat}:
\begin{align}
T_{\beta_{\text{out}},\beta_{\text{in}}}^{W_{\text{out}},W_{\text{in}}}=\sum_{ j_{\sigma}} \dfrac{\omega\tilde \alpha_{W_{\text{out}},\beta_{\text{out}}}^{j_{\sigma}} }{2 (\omega_{j_{\sigma}}-\omega)} \int d \vec r' \vec E_{j_{\sigma}}^{\sigma} \vec W_{\beta_{\text{in}}}  \Delta \varepsilon(\vec r'). 
\end{align}           
Here  $\sigma$ encodes a particular irreducible representation and $j_{\sigma}$ denotes the number of the eigenmode within this representation, $\tilde \alpha_{W_{\text{out}},p_{\text{out}}}^{j_{\sigma}}$ is the multipolar expansion amplitude of the eigenmode $\vec E^{\sigma}_{j_{\sigma}}$, $\Delta \varepsilon(\vec r')= \varepsilon(\vec r')-1$, where $\varepsilon(\vec r)$ is the permittivity function of the resonator,   $\omega_{j_{\sigma}}$ is the complex eigenfrequency. One can see, that all the modes from the same irreducible representation are excited simultaneously but with different efficiency defined, in particular, by proximity of $\omega$ to $\omega_{j_{\sigma}}$.

To conclude, we generalized the concept of bianisotropy beyond the dipole approximation making it applicable to the particles of arbitrary shape and size. We showed that the origin of bianisotropy is absence of the inversion symmetry that results in the coupling between the multipoles of different parity within a single mode. The selection rules for the multipolar coupling are revealed from the group symmetry analysis. The exact analytical expressions defining the strength of the bianisotropic response are derived on the base of the resonant state expansion and T-matrix formalism. The developed theory describes well the optical response even for complicated shapes of the particles when the classical dipole-dipole bianisotropy is violated.  It also predicts new bianisotropy, when either magnetic or electric multipoles are coupled. We believe that the obtained results  are  an important step towards a deeper insight into the scattering properties of nanoantennas and meta-atoms and their smart engineering.

{\it Acknowledgments}
We thank Maxim Gorlach, Daria Smirnova, and Irina Volkovskaya for valuable discussions. The theoretical analysis were performed with supported of the academic leadership program "Priority-2030" and BASIS foundation. The analysis of the bianisotropic response was  supported by RSF (22-42-04420). The Authors thank ITMO University for providing the great research atmosphere.



\bibliography{sample-2}

\begin{thebibliography}{54}%
\makeatletter
\providecommand \@ifxundefined [1]{%
 \@ifx{#1\undefined}
}%
\providecommand \@ifnum [1]{%
 \ifnum #1\expandafter \@firstoftwo
 \else \expandafter \@secondoftwo
 \fi
}%
\providecommand \@ifx [1]{%
 \ifx #1\expandafter \@firstoftwo
 \else \expandafter \@secondoftwo
 \fi
}%
\providecommand \natexlab [1]{#1}%
\providecommand \enquote  [1]{``#1''}%
\providecommand \bibnamefont  [1]{#1}%
\providecommand \bibfnamefont [1]{#1}%
\providecommand \citenamefont [1]{#1}%
\providecommand \href@noop [0]{\@secondoftwo}%
\providecommand \href [0]{\begingroup \@sanitize@url \@href}%
\providecommand \@href[1]{\@@startlink{#1}\@@href}%
\providecommand \@@href[1]{\endgroup#1\@@endlink}%
\providecommand \@sanitize@url [0]{\catcode `\\12\catcode `\$12\catcode
  `\&12\catcode `\#12\catcode `\^12\catcode `\_12\catcode `\%12\relax}%
\providecommand \@@startlink[1]{}%
\providecommand \@@endlink[0]{}%
\providecommand \url  [0]{\begingroup\@sanitize@url \@url }%
\providecommand \@url [1]{\endgroup\@href {#1}{\urlprefix }}%
\providecommand \urlprefix  [0]{URL }%
\providecommand \Eprint [0]{\href }%
\providecommand \doibase [0]{https://doi.org/}%
\providecommand \selectlanguage [0]{\@gobble}%
\providecommand \bibinfo  [0]{\@secondoftwo}%
\providecommand \bibfield  [0]{\@secondoftwo}%
\providecommand \translation [1]{[#1]}%
\providecommand \BibitemOpen [0]{}%
\providecommand \bibitemStop [0]{}%
\providecommand \bibitemNoStop [0]{.\EOS\space}%
\providecommand \EOS [0]{\spacefactor3000\relax}%
\providecommand \BibitemShut  [1]{\csname bibitem#1\endcsname}%
\let\auto@bib@innerbib\@empty
\bibitem [{\citenamefont {Jung}\ \emph {et~al.}(2016)\citenamefont {Jung},
  \citenamefont {In}, \citenamefont {Choi},\ and\ \citenamefont
  {Lee}}]{jung2016electromagnetically}%
  \BibitemOpen
  \bibfield  {author} {\bibinfo {author} {\bibfnamefont {H.}~\bibnamefont
  {Jung}}, \bibinfo {author} {\bibfnamefont {C.}~\bibnamefont {In}}, \bibinfo
  {author} {\bibfnamefont {H.}~\bibnamefont {Choi}},\ and\ \bibinfo {author}
  {\bibfnamefont {H.}~\bibnamefont {Lee}},\ }\bibfield  {title} {\bibinfo
  {title} {Electromagnetically induced transparency analogue by
  self-complementary terahertz meta-atom},\ }\href@noop {} {\bibfield
  {journal} {\bibinfo  {journal} {Adv. Opt. Mater.}\ }\textbf {\bibinfo
  {volume} {4}},\ \bibinfo {pages} {627} (\bibinfo {year} {2016})}\BibitemShut
  {NoStop}%
\bibitem [{\citenamefont {Terekhov}\ \emph {et~al.}(2017)\citenamefont
  {Terekhov}, \citenamefont {Baryshnikova}, \citenamefont {Artemyev},
  \citenamefont {Karabchevsky}, \citenamefont {Shalin},\ and\ \citenamefont
  {Evlyukhin}}]{terekhov2017multipolar}%
  \BibitemOpen
  \bibfield  {author} {\bibinfo {author} {\bibfnamefont {P.~D.}\ \bibnamefont
  {Terekhov}}, \bibinfo {author} {\bibfnamefont {K.~V.}\ \bibnamefont
  {Baryshnikova}}, \bibinfo {author} {\bibfnamefont {Y.~A.}\ \bibnamefont
  {Artemyev}}, \bibinfo {author} {\bibfnamefont {A.}~\bibnamefont
  {Karabchevsky}}, \bibinfo {author} {\bibfnamefont {A.~S.}\ \bibnamefont
  {Shalin}},\ and\ \bibinfo {author} {\bibfnamefont {A.~B.}\ \bibnamefont
  {Evlyukhin}},\ }\bibfield  {title} {\bibinfo {title} {Multipolar response of
  nonspherical silicon nanoparticles in the visible and near-infrared spectral
  ranges},\ }\href@noop {} {\bibfield  {journal} {\bibinfo  {journal} {Phys.
  Rev. B}\ }\textbf {\bibinfo {volume} {96}},\ \bibinfo {pages} {035443}
  (\bibinfo {year} {2017})}\BibitemShut {NoStop}%
\bibitem [{\citenamefont {Koshelev}\ and\ \citenamefont
  {Kivshar}(2020)}]{koshelev2020dielectric}%
  \BibitemOpen
  \bibfield  {author} {\bibinfo {author} {\bibfnamefont {K.}~\bibnamefont
  {Koshelev}}\ and\ \bibinfo {author} {\bibfnamefont {Y.}~\bibnamefont
  {Kivshar}},\ }\bibfield  {title} {\bibinfo {title} {Dielectric resonant
  metaphotonics},\ }\href@noop {} {\bibfield  {journal} {\bibinfo  {journal}
  {ACS Photonics}\ }\textbf {\bibinfo {volume} {8}},\ \bibinfo {pages} {102}
  (\bibinfo {year} {2020})}\BibitemShut {NoStop}%
\bibitem [{\citenamefont {Alaee}\ \emph {et~al.}(2019)\citenamefont {Alaee},
  \citenamefont {Rockstuhl},\ and\ \citenamefont
  {Fernandez-Corbaton}}]{Alaee2019Jan}%
  \BibitemOpen
  \bibfield  {author} {\bibinfo {author} {\bibfnamefont {R.}~\bibnamefont
  {Alaee}}, \bibinfo {author} {\bibfnamefont {C.}~\bibnamefont {Rockstuhl}},\
  and\ \bibinfo {author} {\bibfnamefont {I.}~\bibnamefont
  {Fernandez-Corbaton}},\ }\bibfield  {title} {\bibinfo {title} {{Exact
  Multipolar Decompositions with Applications in Nanophotonics}},\ }\href
  {https://doi.org/10.1002/adom.201800783} {\bibfield  {journal} {\bibinfo
  {journal} {Adv. Opt. Mater.}\ }\textbf {\bibinfo {volume} {7}},\ \bibinfo
  {pages} {1800783} (\bibinfo {year} {2019})}\BibitemShut {NoStop}%
\bibitem [{\citenamefont {Waterman}\ and\ \citenamefont
  {Truell}(1961)}]{waterman1961multiple}%
  \BibitemOpen
  \bibfield  {author} {\bibinfo {author} {\bibfnamefont {P.~C.}\ \bibnamefont
  {Waterman}}\ and\ \bibinfo {author} {\bibfnamefont {R.}~\bibnamefont
  {Truell}},\ }\bibfield  {title} {\bibinfo {title} {Multiple scattering of
  waves},\ }\href@noop {} {\bibfield  {journal} {\bibinfo  {journal} {J. Math.
  Phys.}\ }\textbf {\bibinfo {volume} {2}},\ \bibinfo {pages} {512} (\bibinfo
  {year} {1961})}\BibitemShut {NoStop}%
\bibitem [{\citenamefont {Evlyukhin}\ \emph {et~al.}(2010)\citenamefont
  {Evlyukhin}, \citenamefont {Reinhardt}, \citenamefont {Seidel}, \citenamefont
  {Luk’yanchuk},\ and\ \citenamefont {Chichkov}}]{evlyukhin2010optical}%
  \BibitemOpen
  \bibfield  {author} {\bibinfo {author} {\bibfnamefont {A.~B.}\ \bibnamefont
  {Evlyukhin}}, \bibinfo {author} {\bibfnamefont {C.}~\bibnamefont
  {Reinhardt}}, \bibinfo {author} {\bibfnamefont {A.}~\bibnamefont {Seidel}},
  \bibinfo {author} {\bibfnamefont {B.~S.}\ \bibnamefont {Luk’yanchuk}},\
  and\ \bibinfo {author} {\bibfnamefont {B.~N.}\ \bibnamefont {Chichkov}},\
  }\bibfield  {title} {\bibinfo {title} {Optical response features of
  si-nanoparticle arrays},\ }\href@noop {} {\bibfield  {journal} {\bibinfo
  {journal} {Phys. Rev. B}\ }\textbf {\bibinfo {volume} {82}},\ \bibinfo
  {pages} {045404} (\bibinfo {year} {2010})}\BibitemShut {NoStop}%
\bibitem [{\citenamefont {Miroshnichenko}\ \emph {et~al.}(2015)\citenamefont
  {Miroshnichenko}, \citenamefont {Evlyukhin}, \citenamefont {Yu},
  \citenamefont {Bakker}, \citenamefont {Chipouline}, \citenamefont
  {Kuznetsov}, \citenamefont {Luk’yanchuk}, \citenamefont {Chichkov},\ and\
  \citenamefont {Kivshar}}]{miroshnichenko2015nonradiating}%
  \BibitemOpen
  \bibfield  {author} {\bibinfo {author} {\bibfnamefont {A.~E.}\ \bibnamefont
  {Miroshnichenko}}, \bibinfo {author} {\bibfnamefont {A.~B.}\ \bibnamefont
  {Evlyukhin}}, \bibinfo {author} {\bibfnamefont {Y.~F.}\ \bibnamefont {Yu}},
  \bibinfo {author} {\bibfnamefont {R.~M.}\ \bibnamefont {Bakker}}, \bibinfo
  {author} {\bibfnamefont {A.}~\bibnamefont {Chipouline}}, \bibinfo {author}
  {\bibfnamefont {A.~I.}\ \bibnamefont {Kuznetsov}}, \bibinfo {author}
  {\bibfnamefont {B.}~\bibnamefont {Luk’yanchuk}}, \bibinfo {author}
  {\bibfnamefont {B.~N.}\ \bibnamefont {Chichkov}},\ and\ \bibinfo {author}
  {\bibfnamefont {Y.~S.}\ \bibnamefont {Kivshar}},\ }\bibfield  {title}
  {\bibinfo {title} {Nonradiating anapole modes in dielectric nanoparticles},\
  }\href@noop {} {\bibfield  {journal} {\bibinfo  {journal} {Nat. Commun.}\
  }\textbf {\bibinfo {volume} {6}},\ \bibinfo {pages} {1} (\bibinfo {year}
  {2015})}\BibitemShut {NoStop}%
\bibitem [{\citenamefont {Zenin}\ \emph {et~al.}(2017)\citenamefont {Zenin},
  \citenamefont {Evlyukhin}, \citenamefont {Novikov}, \citenamefont {Yang},
  \citenamefont {Malureanu}, \citenamefont {Lavrinenko}, \citenamefont
  {Chichkov},\ and\ \citenamefont {Bozhevolnyi}}]{zenin2017direct}%
  \BibitemOpen
  \bibfield  {author} {\bibinfo {author} {\bibfnamefont {V.~A.}\ \bibnamefont
  {Zenin}}, \bibinfo {author} {\bibfnamefont {A.~B.}\ \bibnamefont
  {Evlyukhin}}, \bibinfo {author} {\bibfnamefont {S.~M.}\ \bibnamefont
  {Novikov}}, \bibinfo {author} {\bibfnamefont {Y.}~\bibnamefont {Yang}},
  \bibinfo {author} {\bibfnamefont {R.}~\bibnamefont {Malureanu}}, \bibinfo
  {author} {\bibfnamefont {A.~V.}\ \bibnamefont {Lavrinenko}}, \bibinfo
  {author} {\bibfnamefont {B.~N.}\ \bibnamefont {Chichkov}},\ and\ \bibinfo
  {author} {\bibfnamefont {S.~I.}\ \bibnamefont {Bozhevolnyi}},\ }\bibfield
  {title} {\bibinfo {title} {Direct amplitude-phase near-field observation of
  higher-order anapole states},\ }\href@noop {} {\bibfield  {journal} {\bibinfo
   {journal} {Nano Lett.}\ }\textbf {\bibinfo {volume} {17}},\ \bibinfo {pages}
  {7152} (\bibinfo {year} {2017})}\BibitemShut {NoStop}%
\bibitem [{\citenamefont {Baryshnikova}\ \emph {et~al.}(2019)\citenamefont
  {Baryshnikova}, \citenamefont {Smirnova}, \citenamefont {Luk'yanchuk},\ and\
  \citenamefont {Kivshar}}]{baryshnikova2019optical}%
  \BibitemOpen
  \bibfield  {author} {\bibinfo {author} {\bibfnamefont {K.~V.}\ \bibnamefont
  {Baryshnikova}}, \bibinfo {author} {\bibfnamefont {D.~A.}\ \bibnamefont
  {Smirnova}}, \bibinfo {author} {\bibfnamefont {B.~S.}\ \bibnamefont
  {Luk'yanchuk}},\ and\ \bibinfo {author} {\bibfnamefont {Y.~S.}\ \bibnamefont
  {Kivshar}},\ }\bibfield  {title} {\bibinfo {title} {Optical anapoles:
  concepts and applications},\ }\href@noop {} {\bibfield  {journal} {\bibinfo
  {journal} {Adv. Opt. Mater.}\ }\textbf {\bibinfo {volume} {7}},\ \bibinfo
  {pages} {1801350} (\bibinfo {year} {2019})}\BibitemShut {NoStop}%
\bibitem [{\citenamefont {Yang}\ and\ \citenamefont
  {Bozhevolnyi}(2019)}]{yang2019nonradiating}%
  \BibitemOpen
  \bibfield  {author} {\bibinfo {author} {\bibfnamefont {Y.}~\bibnamefont
  {Yang}}\ and\ \bibinfo {author} {\bibfnamefont {S.~I.}\ \bibnamefont
  {Bozhevolnyi}},\ }\bibfield  {title} {\bibinfo {title} {Nonradiating anapole
  states in nanophotonics: from fundamentals to applications},\ }\href@noop {}
  {\bibfield  {journal} {\bibinfo  {journal} {Acs. Sym. Ser.}\ }\textbf
  {\bibinfo {volume} {30}},\ \bibinfo {pages} {204001} (\bibinfo {year}
  {2019})}\BibitemShut {NoStop}%
\bibitem [{\citenamefont {Savinov}\ \emph {et~al.}(2019)\citenamefont
  {Savinov}, \citenamefont {Papasimakis}, \citenamefont {Tsai},\ and\
  \citenamefont {Zheludev}}]{savinov2019optical}%
  \BibitemOpen
  \bibfield  {author} {\bibinfo {author} {\bibfnamefont {V.}~\bibnamefont
  {Savinov}}, \bibinfo {author} {\bibfnamefont {N.}~\bibnamefont
  {Papasimakis}}, \bibinfo {author} {\bibfnamefont {D.}~\bibnamefont {Tsai}},\
  and\ \bibinfo {author} {\bibfnamefont {N.}~\bibnamefont {Zheludev}},\
  }\bibfield  {title} {\bibinfo {title} {Optical anapoles},\ }\href@noop {}
  {\bibfield  {journal} {\bibinfo  {journal} {Commun. Phys.}\ }\textbf
  {\bibinfo {volume} {2}},\ \bibinfo {pages} {1} (\bibinfo {year}
  {2019})}\BibitemShut {NoStop}%
\bibitem [{\citenamefont {Liu}\ and\ \citenamefont
  {Kivshar}(2018)}]{liu2018generalized}%
  \BibitemOpen
  \bibfield  {author} {\bibinfo {author} {\bibfnamefont {W.}~\bibnamefont
  {Liu}}\ and\ \bibinfo {author} {\bibfnamefont {Y.~S.}\ \bibnamefont
  {Kivshar}},\ }\bibfield  {title} {\bibinfo {title} {Generalized kerker
  effects in nanophotonics and meta-optics},\ }\href@noop {} {\bibfield
  {journal} {\bibinfo  {journal} {Opt. Express}\ }\textbf {\bibinfo {volume}
  {26}},\ \bibinfo {pages} {13085} (\bibinfo {year} {2018})}\BibitemShut
  {NoStop}%
\bibitem [{\citenamefont {Shamkhi}\ \emph {et~al.}(2019)\citenamefont
  {Shamkhi}, \citenamefont {Baryshnikova}, \citenamefont {Sayanskiy},
  \citenamefont {Kapitanova}, \citenamefont {Terekhov}, \citenamefont {Belov},
  \citenamefont {Karabchevsky}, \citenamefont {Evlyukhin}, \citenamefont
  {Kivshar},\ and\ \citenamefont {Shalin}}]{shamkhi2019transverse}%
  \BibitemOpen
  \bibfield  {author} {\bibinfo {author} {\bibfnamefont {H.~K.}\ \bibnamefont
  {Shamkhi}}, \bibinfo {author} {\bibfnamefont {K.~V.}\ \bibnamefont
  {Baryshnikova}}, \bibinfo {author} {\bibfnamefont {A.}~\bibnamefont
  {Sayanskiy}}, \bibinfo {author} {\bibfnamefont {P.}~\bibnamefont
  {Kapitanova}}, \bibinfo {author} {\bibfnamefont {P.~D.}\ \bibnamefont
  {Terekhov}}, \bibinfo {author} {\bibfnamefont {P.}~\bibnamefont {Belov}},
  \bibinfo {author} {\bibfnamefont {A.}~\bibnamefont {Karabchevsky}}, \bibinfo
  {author} {\bibfnamefont {A.~B.}\ \bibnamefont {Evlyukhin}}, \bibinfo {author}
  {\bibfnamefont {Y.}~\bibnamefont {Kivshar}},\ and\ \bibinfo {author}
  {\bibfnamefont {A.~S.}\ \bibnamefont {Shalin}},\ }\bibfield  {title}
  {\bibinfo {title} {Transverse scattering and generalized kerker effects in
  all-dielectric mie-resonant metaoptics},\ }\href@noop {} {\bibfield
  {journal} {\bibinfo  {journal} {Phys. Rev. Lett.}\ }\textbf {\bibinfo
  {volume} {122}},\ \bibinfo {pages} {193905} (\bibinfo {year}
  {2019})}\BibitemShut {NoStop}%
\bibitem [{\citenamefont {Fleury}\ \emph {et~al.}(2015)\citenamefont {Fleury},
  \citenamefont {Monticone},\ and\ \citenamefont
  {Al{\`u}}}]{fleury2015invisibility}%
  \BibitemOpen
  \bibfield  {author} {\bibinfo {author} {\bibfnamefont {R.}~\bibnamefont
  {Fleury}}, \bibinfo {author} {\bibfnamefont {F.}~\bibnamefont {Monticone}},\
  and\ \bibinfo {author} {\bibfnamefont {A.}~\bibnamefont {Al{\`u}}},\
  }\bibfield  {title} {\bibinfo {title} {Invisibility and cloaking: Origins,
  present, and future perspectives},\ }\href@noop {} {\bibfield  {journal}
  {\bibinfo  {journal} {Phys. Rev. Appl.}\ }\textbf {\bibinfo {volume} {4}},\
  \bibinfo {pages} {037001} (\bibinfo {year} {2015})}\BibitemShut {NoStop}%
\bibitem [{\citenamefont {Babicheva}\ and\ \citenamefont
  {Evlyukhin}(2021)}]{babicheva2021multipole}%
  \BibitemOpen
  \bibfield  {author} {\bibinfo {author} {\bibfnamefont {V.~E.}\ \bibnamefont
  {Babicheva}}\ and\ \bibinfo {author} {\bibfnamefont {A.~B.}\ \bibnamefont
  {Evlyukhin}},\ }\bibfield  {title} {\bibinfo {title} {Multipole lattice
  effects in high refractive index metasurfaces},\ }\href@noop {} {\bibfield
  {journal} {\bibinfo  {journal} {J. Appl. Phys.}\ }\textbf {\bibinfo {volume}
  {129}},\ \bibinfo {pages} {040902} (\bibinfo {year} {2021})}\BibitemShut
  {NoStop}%
\bibitem [{\citenamefont {Alaee}\ \emph {et~al.}(2015)\citenamefont {Alaee},
  \citenamefont {Albooyeh}, \citenamefont {Rahimzadegan}, \citenamefont
  {Mirmoosa}, \citenamefont {Kivshar},\ and\ \citenamefont
  {Rockstuhl}}]{Alaee2015-All-dielectricrecipr}%
  \BibitemOpen
  \bibfield  {author} {\bibinfo {author} {\bibfnamefont {R.}~\bibnamefont
  {Alaee}}, \bibinfo {author} {\bibfnamefont {M.}~\bibnamefont {Albooyeh}},
  \bibinfo {author} {\bibfnamefont {A.}~\bibnamefont {Rahimzadegan}}, \bibinfo
  {author} {\bibfnamefont {M.~S.}\ \bibnamefont {Mirmoosa}}, \bibinfo {author}
  {\bibfnamefont {Y.~S.}\ \bibnamefont {Kivshar}},\ and\ \bibinfo {author}
  {\bibfnamefont {C.}~\bibnamefont {Rockstuhl}},\ }\bibfield  {title} {\bibinfo
  {title} {{All-dielectric reciprocal bianisotropic nanoparticles}},\ }\href
  {https://doi.org/10.1103/PhysRevB.92.245130} {\bibfield  {journal} {\bibinfo
  {journal} {Phys. Rev. B}\ }\textbf {\bibinfo {volume} {92}},\ \bibinfo
  {pages} {245130} (\bibinfo {year} {2015})}\BibitemShut {NoStop}%
\bibitem [{\citenamefont {Bobylev}\ \emph {et~al.}(2020)\citenamefont
  {Bobylev}, \citenamefont {Smirnova},\ and\ \citenamefont
  {Gorlach}}]{Bobylev2020-Nonlocalresponseof}%
  \BibitemOpen
  \bibfield  {author} {\bibinfo {author} {\bibfnamefont {D.~A.}\ \bibnamefont
  {Bobylev}}, \bibinfo {author} {\bibfnamefont {D.~A.}\ \bibnamefont
  {Smirnova}},\ and\ \bibinfo {author} {\bibfnamefont {M.~A.}\ \bibnamefont
  {Gorlach}},\ }\bibfield  {title} {\bibinfo {title} {{Nonlocal response of
  Mie-resonant dielectric particles}},\ }\href
  {https://doi.org/10.1103/PhysRevB.102.115110} {\bibfield  {journal} {\bibinfo
   {journal} {Phys. Rev. B}\ }\textbf {\bibinfo {volume} {102}},\ \bibinfo
  {pages} {115110} (\bibinfo {year} {2020})}\BibitemShut {NoStop}%
\bibitem [{\citenamefont {Evlyukhin}\ \emph {et~al.}(2020)\citenamefont
  {Evlyukhin}, \citenamefont {Tuz}, \citenamefont {Volkov},\ and\ \citenamefont
  {Chichkov}}]{evlyukhin2020bianisotropy}%
  \BibitemOpen
  \bibfield  {author} {\bibinfo {author} {\bibfnamefont {A.~B.}\ \bibnamefont
  {Evlyukhin}}, \bibinfo {author} {\bibfnamefont {V.~R.}\ \bibnamefont {Tuz}},
  \bibinfo {author} {\bibfnamefont {V.~S.}\ \bibnamefont {Volkov}},\ and\
  \bibinfo {author} {\bibfnamefont {B.~N.}\ \bibnamefont {Chichkov}},\
  }\bibfield  {title} {\bibinfo {title} {Bianisotropy for light trapping in
  all-dielectric metasurfaces},\ }\href
  {https://doi.org/10.1103/PhysRevB.101.205415} {\bibfield  {journal} {\bibinfo
   {journal} {Phys. Rev. B}\ }\textbf {\bibinfo {volume} {101}},\ \bibinfo
  {pages} {205415} (\bibinfo {year} {2020})}\BibitemShut {NoStop}%
\bibitem [{\citenamefont {Albooyeh}\ \emph {et~al.}(2016)\citenamefont
  {Albooyeh}, \citenamefont {Asadchy}, \citenamefont {Alaee}, \citenamefont
  {Hashemi}, \citenamefont {Yazdi}, \citenamefont {Mirmoosa}, \citenamefont
  {Rockstuhl}, \citenamefont {Simovski},\ and\ \citenamefont
  {Tretyakov}}]{Albooyeh2016Dec}%
  \BibitemOpen
  \bibfield  {author} {\bibinfo {author} {\bibfnamefont {M.}~\bibnamefont
  {Albooyeh}}, \bibinfo {author} {\bibfnamefont {V.~S.}\ \bibnamefont
  {Asadchy}}, \bibinfo {author} {\bibfnamefont {R.}~\bibnamefont {Alaee}},
  \bibinfo {author} {\bibfnamefont {S.~M.}\ \bibnamefont {Hashemi}}, \bibinfo
  {author} {\bibfnamefont {M.}~\bibnamefont {Yazdi}}, \bibinfo {author}
  {\bibfnamefont {M.~S.}\ \bibnamefont {Mirmoosa}}, \bibinfo {author}
  {\bibfnamefont {C.}~\bibnamefont {Rockstuhl}}, \bibinfo {author}
  {\bibfnamefont {C.~R.}\ \bibnamefont {Simovski}},\ and\ \bibinfo {author}
  {\bibfnamefont {S.~A.}\ \bibnamefont {Tretyakov}},\ }\bibfield  {title}
  {\bibinfo {title} {{Purely bianisotropic scatterers}},\ }\href
  {https://doi.org/10.1103/PhysRevB.94.245428} {\bibfield  {journal} {\bibinfo
  {journal} {Phys. Rev. B}\ }\textbf {\bibinfo {volume} {94}},\ \bibinfo
  {pages} {245428} (\bibinfo {year} {2016})}\BibitemShut {NoStop}%
\bibitem [{\citenamefont {Asadchy}\ \emph {et~al.}(2018)\citenamefont
  {Asadchy}, \citenamefont {D{\'\i}az-Rubio},\ and\ \citenamefont
  {Tretyakov}}]{asadchy2018bianisotropic}%
  \BibitemOpen
  \bibfield  {author} {\bibinfo {author} {\bibfnamefont {V.~S.}\ \bibnamefont
  {Asadchy}}, \bibinfo {author} {\bibfnamefont {A.}~\bibnamefont
  {D{\'\i}az-Rubio}},\ and\ \bibinfo {author} {\bibfnamefont {S.~A.}\
  \bibnamefont {Tretyakov}},\ }\bibfield  {title} {\bibinfo {title}
  {Bianisotropic metasurfaces: physics and applications},\ }\href@noop {}
  {\bibfield  {journal} {\bibinfo  {journal} {P. Soc. Photo-opt. Ins.}\
  }\textbf {\bibinfo {volume} {7}},\ \bibinfo {pages} {1069} (\bibinfo {year}
  {2018})}\BibitemShut {NoStop}%
\bibitem [{\citenamefont {Rodriguez}\ \emph {et~al.}(2014)\citenamefont
  {Rodriguez}, \citenamefont {Arango}, \citenamefont {Steinbusch},
  \citenamefont {Verschuuren}, \citenamefont {Koenderink},\ and\ \citenamefont
  {Rivas}}]{Rodriguez2014Dec}%
  \BibitemOpen
  \bibfield  {author} {\bibinfo {author} {\bibfnamefont {S.~R.~K.}\
  \bibnamefont {Rodriguez}}, \bibinfo {author} {\bibfnamefont {F.~B.}\
  \bibnamefont {Arango}}, \bibinfo {author} {\bibfnamefont {T.~P.}\
  \bibnamefont {Steinbusch}}, \bibinfo {author} {\bibfnamefont {M.~A.}\
  \bibnamefont {Verschuuren}}, \bibinfo {author} {\bibfnamefont {A.~F.}\
  \bibnamefont {Koenderink}},\ and\ \bibinfo {author} {\bibfnamefont {J.~G.}\
  \bibnamefont {Rivas}},\ }\bibfield  {title} {\bibinfo {title} {{Breaking the
  Symmetry of Forward-Backward Light Emission with Localized and Collective
  Magnetoelectric Resonances in Arrays of Pyramid-Shaped Aluminum
  Nanoparticles}},\ }\href {https://doi.org/10.1103/PhysRevLett.113.247401}
  {\bibfield  {journal} {\bibinfo  {journal} {Phys. Rev. Lett.}\ }\textbf
  {\bibinfo {volume} {113}},\ \bibinfo {pages} {247401} (\bibinfo {year}
  {2014})}\BibitemShut {NoStop}%
\bibitem [{\citenamefont {Achouri}\ \emph {et~al.}(2020)\citenamefont
  {Achouri}, \citenamefont {Kiselev},\ and\ \citenamefont
  {Martin}}]{Achouri2020Aug}%
  \BibitemOpen
  \bibfield  {author} {\bibinfo {author} {\bibfnamefont {K.}~\bibnamefont
  {Achouri}}, \bibinfo {author} {\bibfnamefont {A.}~\bibnamefont {Kiselev}},\
  and\ \bibinfo {author} {\bibfnamefont {O.~J.~F.}\ \bibnamefont {Martin}},\
  }\bibfield  {title} {\bibinfo {title} {{Multipolar origin of electromagnetic
  transverse force resulting from two-wave interference}},\ }\href
  {https://doi.org/10.1103/PhysRevB.102.085107} {\bibfield  {journal} {\bibinfo
   {journal} {Phys. Rev. B}\ }\textbf {\bibinfo {volume} {102}},\ \bibinfo
  {pages} {085107} (\bibinfo {year} {2020})}\BibitemShut {NoStop}%
\bibitem [{\citenamefont {Wei}\ and\ \citenamefont
  {Rodr{\ifmmode\acute{\imath}\else\'{\i}\fi}guez-Fortu{\ifmmode\tilde{n}\else\~{n}\fi}o}(2022)}]{Wei2022Mar}%
  \BibitemOpen
  \bibfield  {author} {\bibinfo {author} {\bibfnamefont {L.}~\bibnamefont
  {Wei}}\ and\ \bibinfo {author} {\bibfnamefont {F.~J.}\ \bibnamefont
  {Rodr{\ifmmode\acute{\imath}\else\'{\i}\fi}guez-Fortu{\ifmmode\tilde{n}\else\~{n}\fi}o}},\
  }\bibfield  {title} {\bibinfo {title} {{Optical multipolar torque in
  structured electromagnetic fields: On helicity gradient torque, quadrupolar
  torque, and spin of the field gradient}},\ }\href
  {https://doi.org/10.1103/PhysRevB.105.125424} {\bibfield  {journal} {\bibinfo
   {journal} {Phys. Rev. B}\ }\textbf {\bibinfo {volume} {105}},\ \bibinfo
  {pages} {125424} (\bibinfo {year} {2022})}\BibitemShut {NoStop}%
\bibitem [{\citenamefont {Wu}\ \emph {et~al.}(2020)\citenamefont {Wu},
  \citenamefont {Baron}, \citenamefont {Lalanne},\ and\ \citenamefont
  {Vynck}}]{Wu2020Jan}%
  \BibitemOpen
  \bibfield  {author} {\bibinfo {author} {\bibfnamefont {T.}~\bibnamefont
  {Wu}}, \bibinfo {author} {\bibfnamefont {A.}~\bibnamefont {Baron}}, \bibinfo
  {author} {\bibfnamefont {P.}~\bibnamefont {Lalanne}},\ and\ \bibinfo {author}
  {\bibfnamefont {K.}~\bibnamefont {Vynck}},\ }\bibfield  {title} {\bibinfo
  {title} {{Intrinsic multipolar contents of nanoresonators for tailored
  scattering}},\ }\href {https://doi.org/10.1103/PhysRevA.101.011803}
  {\bibfield  {journal} {\bibinfo  {journal} {Phys. Rev. A}\ }\textbf {\bibinfo
  {volume} {101}},\ \bibinfo {pages} {011803(R)} (\bibinfo {year}
  {2020})}\BibitemShut {NoStop}%
\bibitem [{\citenamefont {M{\ifmmode\ddot{u}\else\"{u}\fi}hlig}\ \emph
  {et~al.}(2011)\citenamefont {M{\ifmmode\ddot{u}\else\"{u}\fi}hlig},
  \citenamefont {Menzel}, \citenamefont {Rockstuhl},\ and\ \citenamefont
  {Lederer}}]{Muhlig2011Jun}%
  \BibitemOpen
  \bibfield  {author} {\bibinfo {author} {\bibfnamefont {S.}~\bibnamefont
  {M{\ifmmode\ddot{u}\else\"{u}\fi}hlig}}, \bibinfo {author} {\bibfnamefont
  {C.}~\bibnamefont {Menzel}}, \bibinfo {author} {\bibfnamefont
  {C.}~\bibnamefont {Rockstuhl}},\ and\ \bibinfo {author} {\bibfnamefont
  {F.}~\bibnamefont {Lederer}},\ }\bibfield  {title} {\bibinfo {title}
  {{Multipole analysis of meta-atoms}},\ }\href
  {https://doi.org/10.1016/j.metmat.2011.03.003} {\bibfield  {journal}
  {\bibinfo  {journal} {Metamaterials}\ }\textbf {\bibinfo {volume} {5}},\
  \bibinfo {pages} {64} (\bibinfo {year} {2011})}\BibitemShut {NoStop}%
\bibitem [{\citenamefont {Gladyshev}\ \emph {et~al.}(2020)\citenamefont
  {Gladyshev}, \citenamefont {Frizyuk},\ and\ \citenamefont
  {Bogdanov}}]{Gladyshev2020Aug}%
  \BibitemOpen
  \bibfield  {author} {\bibinfo {author} {\bibfnamefont {S.}~\bibnamefont
  {Gladyshev}}, \bibinfo {author} {\bibfnamefont {K.}~\bibnamefont {Frizyuk}},\
  and\ \bibinfo {author} {\bibfnamefont {A.}~\bibnamefont {Bogdanov}},\
  }\bibfield  {title} {\bibinfo {title} {{Symmetry analysis and multipole
  classification of eigenmodes in electromagnetic resonators for engineering
  their optical properties}},\ }\href
  {https://doi.org/10.1103/PhysRevB.102.075103} {\bibfield  {journal} {\bibinfo
   {journal} {Phys. Rev. B}\ }\textbf {\bibinfo {volume} {102}},\ \bibinfo
  {pages} {075103} (\bibinfo {year} {2020})}\BibitemShut {NoStop}%
\bibitem [{\citenamefont {Alaee}\ \emph {et~al.}(2018)\citenamefont {Alaee},
  \citenamefont {Rockstuhl},\ and\ \citenamefont
  {Fernandez-Corbaton}}]{Alaee2017Jan}%
  \BibitemOpen
  \bibfield  {author} {\bibinfo {author} {\bibfnamefont {R.}~\bibnamefont
  {Alaee}}, \bibinfo {author} {\bibfnamefont {C.}~\bibnamefont {Rockstuhl}},\
  and\ \bibinfo {author} {\bibfnamefont {I.}~\bibnamefont
  {Fernandez-Corbaton}},\ }\bibfield  {title} {\bibinfo {title} {{An
  electromagnetic multipole expansion beyond the long-wavelength
  approximation}},\ }\href {https://doi.org/10.1016/j.optcom.2017.08.064}
  {\bibfield  {journal} {\bibinfo  {journal} {Opt. Commun.}\ }\textbf {\bibinfo
  {volume} {407}},\ \bibinfo {pages} {17} (\bibinfo {year} {2018})}\BibitemShut
  {NoStop}%
\bibitem [{\citenamefont {Grinblat}\ \emph {et~al.}(2017)\citenamefont
  {Grinblat}, \citenamefont {Li}, \citenamefont {Nielsen}, \citenamefont
  {Oulton},\ and\ \citenamefont {Maier}}]{grinblat2017efficient}%
  \BibitemOpen
  \bibfield  {author} {\bibinfo {author} {\bibfnamefont {G.}~\bibnamefont
  {Grinblat}}, \bibinfo {author} {\bibfnamefont {Y.}~\bibnamefont {Li}},
  \bibinfo {author} {\bibfnamefont {M.~P.}\ \bibnamefont {Nielsen}}, \bibinfo
  {author} {\bibfnamefont {R.~F.}\ \bibnamefont {Oulton}},\ and\ \bibinfo
  {author} {\bibfnamefont {S.~A.}\ \bibnamefont {Maier}},\ }\bibfield  {title}
  {\bibinfo {title} {Efficient third harmonic generation and nonlinear
  subwavelength imaging at a higher-order anapole mode in a single germanium
  nanodisk},\ }\href@noop {} {\bibfield  {journal} {\bibinfo  {journal} {ACS
  Nano}\ }\textbf {\bibinfo {volume} {11}},\ \bibinfo {pages} {953} (\bibinfo
  {year} {2017})}\BibitemShut {NoStop}%
\bibitem [{\citenamefont {Bogdanov}\ \emph {et~al.}(2019)\citenamefont
  {Bogdanov}, \citenamefont {Koshelev}, \citenamefont {Kapitanova},
  \citenamefont {Rybin}, \citenamefont {Gladyshev}, \citenamefont {Sadrieva},
  \citenamefont {Samusev}, \citenamefont {Kivshar},\ and\ \citenamefont
  {Limonov}}]{bogdanov2019bound}%
  \BibitemOpen
  \bibfield  {author} {\bibinfo {author} {\bibfnamefont {A.~A.}\ \bibnamefont
  {Bogdanov}}, \bibinfo {author} {\bibfnamefont {K.~L.}\ \bibnamefont
  {Koshelev}}, \bibinfo {author} {\bibfnamefont {P.~V.}\ \bibnamefont
  {Kapitanova}}, \bibinfo {author} {\bibfnamefont {M.~V.}\ \bibnamefont
  {Rybin}}, \bibinfo {author} {\bibfnamefont {S.~A.}\ \bibnamefont
  {Gladyshev}}, \bibinfo {author} {\bibfnamefont {Z.~F.}\ \bibnamefont
  {Sadrieva}}, \bibinfo {author} {\bibfnamefont {K.~B.}\ \bibnamefont
  {Samusev}}, \bibinfo {author} {\bibfnamefont {Y.~S.}\ \bibnamefont
  {Kivshar}},\ and\ \bibinfo {author} {\bibfnamefont {M.~F.}\ \bibnamefont
  {Limonov}},\ }\bibfield  {title} {\bibinfo {title} {Bound states in the
  continuum and fano resonances in the strong mode coupling regime},\
  }\href@noop {} {\bibfield  {journal} {\bibinfo  {journal} {Advanced
  Photonics}\ }\textbf {\bibinfo {volume} {1}},\ \bibinfo {pages} {016001}
  (\bibinfo {year} {2019})}\BibitemShut {NoStop}%
\bibitem [{\citenamefont {Odit}\ \emph {et~al.}(2021)\citenamefont {Odit},
  \citenamefont {Koshelev}, \citenamefont {Gladyshev}, \citenamefont
  {Ladutenko}, \citenamefont {Kivshar},\ and\ \citenamefont
  {Bogdanov}}]{odit2021observation}%
  \BibitemOpen
  \bibfield  {author} {\bibinfo {author} {\bibfnamefont {M.}~\bibnamefont
  {Odit}}, \bibinfo {author} {\bibfnamefont {K.}~\bibnamefont {Koshelev}},
  \bibinfo {author} {\bibfnamefont {S.}~\bibnamefont {Gladyshev}}, \bibinfo
  {author} {\bibfnamefont {K.}~\bibnamefont {Ladutenko}}, \bibinfo {author}
  {\bibfnamefont {Y.}~\bibnamefont {Kivshar}},\ and\ \bibinfo {author}
  {\bibfnamefont {A.}~\bibnamefont {Bogdanov}},\ }\bibfield  {title} {\bibinfo
  {title} {Observation of supercavity modes in subwavelength dielectric
  resonators},\ }\href@noop {} {\bibfield  {journal} {\bibinfo  {journal} {Adv.
  Mater.}\ }\textbf {\bibinfo {volume} {33}},\ \bibinfo {pages} {2003804}
  (\bibinfo {year} {2021})}\BibitemShut {NoStop}%
\bibitem [{\citenamefont {Sakai}\ \emph {et~al.}(2015)\citenamefont {Sakai},
  \citenamefont {Nomura}, \citenamefont {Yamamoto},\ and\ \citenamefont
  {Sasaki}}]{sakai2015excitation}%
  \BibitemOpen
  \bibfield  {author} {\bibinfo {author} {\bibfnamefont {K.}~\bibnamefont
  {Sakai}}, \bibinfo {author} {\bibfnamefont {K.}~\bibnamefont {Nomura}},
  \bibinfo {author} {\bibfnamefont {T.}~\bibnamefont {Yamamoto}},\ and\
  \bibinfo {author} {\bibfnamefont {K.}~\bibnamefont {Sasaki}},\ }\bibfield
  {title} {\bibinfo {title} {Excitation of multipole plasmons by optical vortex
  beams},\ }\href@noop {} {\bibfield  {journal} {\bibinfo  {journal} {Sci.
  Rep.}\ }\textbf {\bibinfo {volume} {5}},\ \bibinfo {pages} {1} (\bibinfo
  {year} {2015})}\BibitemShut {NoStop}%
\bibitem [{\citenamefont {Wo{\'z}niak}\ \emph {et~al.}(2015)\citenamefont
  {Wo{\'z}niak}, \citenamefont {Banzer},\ and\ \citenamefont
  {Leuchs}}]{wozniak2015selective}%
  \BibitemOpen
  \bibfield  {author} {\bibinfo {author} {\bibfnamefont {P.}~\bibnamefont
  {Wo{\'z}niak}}, \bibinfo {author} {\bibfnamefont {P.}~\bibnamefont
  {Banzer}},\ and\ \bibinfo {author} {\bibfnamefont {G.}~\bibnamefont
  {Leuchs}},\ }\bibfield  {title} {\bibinfo {title} {Selective switching of
  individual multipole resonances in single dielectric nanoparticles},\
  }\href@noop {} {\bibfield  {journal} {\bibinfo  {journal} {Laser Photonics
  Rev.}\ }\textbf {\bibinfo {volume} {9}},\ \bibinfo {pages} {231} (\bibinfo
  {year} {2015})}\BibitemShut {NoStop}%
\bibitem [{\citenamefont {Das}\ \emph {et~al.}(2015{\natexlab{a}})\citenamefont
  {Das}, \citenamefont {Iyer}, \citenamefont {DeCrescent},\ and\ \citenamefont
  {Schuller}}]{Das2015Dec}%
  \BibitemOpen
  \bibfield  {author} {\bibinfo {author} {\bibfnamefont {T.}~\bibnamefont
  {Das}}, \bibinfo {author} {\bibfnamefont {P.~P.}\ \bibnamefont {Iyer}},
  \bibinfo {author} {\bibfnamefont {R.~A.}\ \bibnamefont {DeCrescent}},\ and\
  \bibinfo {author} {\bibfnamefont {J.~A.}\ \bibnamefont {Schuller}},\
  }\bibfield  {title} {\bibinfo {title} {{Beam engineering for selective and
  enhanced coupling to multipolar resonances}},\ }\href
  {https://doi.org/10.1103/PhysRevB.92.241110} {\bibfield  {journal} {\bibinfo
  {journal} {Phys. Rev. B}\ }\textbf {\bibinfo {volume} {92}},\ \bibinfo
  {pages} {241110} (\bibinfo {year} {2015}{\natexlab{a}})}\BibitemShut
  {NoStop}%
\bibitem [{\citenamefont {Proust}\ \emph {et~al.}(2016)\citenamefont {Proust},
  \citenamefont {Bonod}, \citenamefont {Grand},\ and\ \citenamefont
  {Gallas}}]{Proust2016Sep}%
  \BibitemOpen
  \bibfield  {author} {\bibinfo {author} {\bibfnamefont {J.}~\bibnamefont
  {Proust}}, \bibinfo {author} {\bibfnamefont {N.}~\bibnamefont {Bonod}},
  \bibinfo {author} {\bibfnamefont {J.}~\bibnamefont {Grand}},\ and\ \bibinfo
  {author} {\bibfnamefont {B.}~\bibnamefont {Gallas}},\ }\bibfield  {title}
  {\bibinfo {title} {{Optical Monitoring of the Magnetoelectric Coupling in
  Individual Plasmonic Scatterers}},\ }\href
  {https://doi.org/10.1021/acsphotonics.6b00041} {\bibfield  {journal}
  {\bibinfo  {journal} {ACS Photonics}\ }\textbf {\bibinfo {volume} {3}},\
  \bibinfo {pages} {1581} (\bibinfo {year} {2016})}\BibitemShut {NoStop}%
\bibitem [{\citenamefont {Poleva}\ \emph {et~al.}(2021)\citenamefont {Poleva},
  \citenamefont {Baryshnikova}, \citenamefont {Frizyuk},\ and\ \citenamefont
  {Evlyukhin}}]{Poleva2021Nov}%
  \BibitemOpen
  \bibfield  {author} {\bibinfo {author} {\bibfnamefont {M.}~\bibnamefont
  {Poleva}}, \bibinfo {author} {\bibfnamefont {K.~V.}\ \bibnamefont
  {Baryshnikova}}, \bibinfo {author} {\bibfnamefont {K.}~\bibnamefont
  {Frizyuk}},\ and\ \bibinfo {author} {\bibfnamefont {A.~B.}\ \bibnamefont
  {Evlyukhin}},\ }\bibfield  {title} {\bibinfo {title} {{Nontrivial optical
  response of silicon triangular prisms}},\ }\href
  {https://doi.org/10.1088/1742-6596/2015/1/012111} {\bibfield  {journal}
  {\bibinfo  {journal} {J. Phys. Conf. Ser.}\ }\textbf {\bibinfo {volume}
  {2015}},\ \bibinfo {pages} {012111} (\bibinfo {year} {2021})}\BibitemShut
  {NoStop}%
\bibitem [{\citenamefont {Cheng}\ and\ \citenamefont
  {Kong}(1968{\natexlab{a}})}]{cheng1968covariant}%
  \BibitemOpen
  \bibfield  {author} {\bibinfo {author} {\bibfnamefont {D.~K.}\ \bibnamefont
  {Cheng}}\ and\ \bibinfo {author} {\bibfnamefont {J.-A.}\ \bibnamefont
  {Kong}},\ }\bibfield  {title} {\bibinfo {title} {Covariant descriptions of
  bianisotropic media},\ }\href@noop {} {\bibfield  {journal} {\bibinfo
  {journal} {Proc. IEEE}\ }\textbf {\bibinfo {volume} {56}},\ \bibinfo {pages}
  {248} (\bibinfo {year} {1968}{\natexlab{a}})}\BibitemShut {NoStop}%
\bibitem [{\citenamefont {Cheng}\ and\ \citenamefont
  {Kong}(1968{\natexlab{b}})}]{cheng1968time}%
  \BibitemOpen
  \bibfield  {author} {\bibinfo {author} {\bibfnamefont {D.~K.}\ \bibnamefont
  {Cheng}}\ and\ \bibinfo {author} {\bibfnamefont {J.-A.}\ \bibnamefont
  {Kong}},\ }\bibfield  {title} {\bibinfo {title} {Time-harmonic fields in
  source-free bianisotropic media},\ }\href@noop {} {\bibfield  {journal}
  {\bibinfo  {journal} {J. Appl. Phys.}\ }\textbf {\bibinfo {volume} {39}},\
  \bibinfo {pages} {5792} (\bibinfo {year} {1968}{\natexlab{b}})}\BibitemShut
  {NoStop}%
\bibitem [{\citenamefont {Fan}\ \emph {et~al.}(2013)\citenamefont {Fan},
  \citenamefont {Strikwerda}, \citenamefont {Zhang},\ and\ \citenamefont
  {Averitt}}]{Fan2013Apr}%
  \BibitemOpen
  \bibfield  {author} {\bibinfo {author} {\bibfnamefont {K.}~\bibnamefont
  {Fan}}, \bibinfo {author} {\bibfnamefont {A.~C.}\ \bibnamefont {Strikwerda}},
  \bibinfo {author} {\bibfnamefont {X.}~\bibnamefont {Zhang}},\ and\ \bibinfo
  {author} {\bibfnamefont {R.~D.}\ \bibnamefont {Averitt}},\ }\bibfield
  {title} {\bibinfo {title} {{Three-dimensional broadband tunable terahertz
  metamaterials}},\ }\href {https://doi.org/10.1103/PhysRevB.87.161104}
  {\bibfield  {journal} {\bibinfo  {journal} {Phys. Rev. B}\ }\textbf {\bibinfo
  {volume} {87}},\ \bibinfo {pages} {161104} (\bibinfo {year}
  {2013})}\BibitemShut {NoStop}%
\bibitem [{\citenamefont {Ivchenko}\ and\ \citenamefont
  {Pikus}(2012)}]{ivchenko2012superlattices}%
  \BibitemOpen
  \bibfield  {author} {\bibinfo {author} {\bibfnamefont {E.~L.}\ \bibnamefont
  {Ivchenko}}\ and\ \bibinfo {author} {\bibfnamefont {G.}~\bibnamefont
  {Pikus}},\ }\href@noop {} {\emph {\bibinfo {title} {Superlattices and other
  heterostructures: symmetry and optical phenomena}}},\ Vol.\ \bibinfo {volume}
  {110}\ (\bibinfo  {publisher} {Springer},\ \bibinfo {address} {Berlin},\
  \bibinfo {year} {2012})\BibitemShut {NoStop}%
\bibitem [{\citenamefont {Xiong}\ \emph {et~al.}(2020)\citenamefont {Xiong},
  \citenamefont {Xiong}, \citenamefont {Yang}, \citenamefont {Yang},
  \citenamefont {Chen}, \citenamefont {Wang}, \citenamefont {Xu}, \citenamefont
  {Xu}, \citenamefont {Xu}, \citenamefont {Liu}, \citenamefont {Chen},
  \citenamefont {Chen},\ and\ \citenamefont {Chen}}]{Xiong2020Feb}%
  \BibitemOpen
  \bibfield  {author} {\bibinfo {author} {\bibfnamefont {Z.}~\bibnamefont
  {Xiong}}, \bibinfo {author} {\bibfnamefont {Z.}~\bibnamefont {Xiong}},
  \bibinfo {author} {\bibfnamefont {Q.}~\bibnamefont {Yang}}, \bibinfo {author}
  {\bibfnamefont {Q.}~\bibnamefont {Yang}}, \bibinfo {author} {\bibfnamefont
  {W.}~\bibnamefont {Chen}}, \bibinfo {author} {\bibfnamefont {Z.}~\bibnamefont
  {Wang}}, \bibinfo {author} {\bibfnamefont {J.}~\bibnamefont {Xu}}, \bibinfo
  {author} {\bibfnamefont {J.}~\bibnamefont {Xu}}, \bibinfo {author}
  {\bibfnamefont {J.}~\bibnamefont {Xu}}, \bibinfo {author} {\bibfnamefont
  {W.}~\bibnamefont {Liu}}, \bibinfo {author} {\bibfnamefont {Y.}~\bibnamefont
  {Chen}}, \bibinfo {author} {\bibfnamefont {Y.}~\bibnamefont {Chen}},\ and\
  \bibinfo {author} {\bibfnamefont {Y.}~\bibnamefont {Chen}},\ }\bibfield
  {title} {\bibinfo {title} {{On the constraints of electromagnetic multipoles
  for symmetric scatterers: eigenmode analysis}},\ }\href
  {https://doi.org/10.1364/OE.382239} {\bibfield  {journal} {\bibinfo
  {journal} {Opt. Express}\ }\textbf {\bibinfo {volume} {28}},\ \bibinfo
  {pages} {3073} (\bibinfo {year} {2020})}\BibitemShut {NoStop}%
\bibitem [{\citenamefont {Tsimokha}\ \emph {et~al.}(2021)\citenamefont
  {Tsimokha}, \citenamefont {Igoshin}, \citenamefont {Nikitina}, \citenamefont
  {Petrov}, \citenamefont {Toftul},\ and\ \citenamefont
  {Frizyuk}}]{Tsimokha2021-Acousticresonators}%
  \BibitemOpen
  \bibfield  {author} {\bibinfo {author} {\bibfnamefont {M.}~\bibnamefont
  {Tsimokha}}, \bibinfo {author} {\bibfnamefont {V.}~\bibnamefont {Igoshin}},
  \bibinfo {author} {\bibfnamefont {A.}~\bibnamefont {Nikitina}}, \bibinfo
  {author} {\bibfnamefont {M.}~\bibnamefont {Petrov}}, \bibinfo {author}
  {\bibfnamefont {I.}~\bibnamefont {Toftul}},\ and\ \bibinfo {author}
  {\bibfnamefont {K.}~\bibnamefont {Frizyuk}},\ }\bibfield  {title} {\bibinfo
  {title} {{Acoustic resonators: symmetry classification and multipolar content
  of the eigenmodes}},\ }\href {https://arxiv.org/abs/2110.11220v1} {\bibfield
  {journal} {\bibinfo  {journal} {arXiv}\ } (\bibinfo {year} {2021})},\ \Eprint
  {https://arxiv.org/abs/2110.11220} {2110.11220} \BibitemShut {NoStop}%
\bibitem [{\citenamefont {Pir{\ifmmode\acute{o}\else\'{o}\fi}th}\ and\
  \citenamefont
  {S{\ifmmode\acute{o}\else\'{o}\fi}lyom}(2007)}]{Piroth2007-FundamentalsoftheP}%
  \BibitemOpen
  \bibfield  {author} {\bibinfo {author} {\bibfnamefont {A.}~\bibnamefont
  {Pir{\ifmmode\acute{o}\else\'{o}\fi}th}}\ and\ \bibinfo {author}
  {\bibfnamefont {J.}~\bibnamefont {S{\ifmmode\acute{o}\else\'{o}\fi}lyom}},\
  }\href
  {https://books.google.ru/books?id=zn-se2TKv3QC&pg=PA172&lpg=PA172&dq=Fundamentals+of+the+Physics+of+Solids:+Volume+1:+Structure+and+Dynamics+wigner+theorem&source=bl&ots=Jnh7vmrIX3&sig=ACfU3U2HDSyHz3GIPeQB8WMsFiBP9ndq4g&hl=ru&sa=X&ved=2ahUKEwiSk9jHno7oAhUxyKYKHQAqCPoQ6AEwAXoECB4QAQ#v=onepage&q=Fundamentals\%20of\%20the\%20Physics\%20of\%20Solids\%3A\%20Volume\%201\%3A\%20Structure\%20and\%20Dynamics\%20wigner\%20theorem&f=false}
  {\emph {\bibinfo {title} {{Fundamentals of the Physics of Solids}}}}\
  (\bibinfo  {publisher} {Springer},\ \bibinfo {address} {Berlin, Germany},\
  \bibinfo {year} {2007})\BibitemShut {NoStop}%
\bibitem [{\citenamefont {Bohren}\ and\ \citenamefont
  {Huffman}(1998)}]{Bohren1998Mar}%
  \BibitemOpen
  \bibfield  {author} {\bibinfo {author} {\bibfnamefont {C.~F.}\ \bibnamefont
  {Bohren}}\ and\ \bibinfo {author} {\bibfnamefont {D.~R.}\ \bibnamefont
  {Huffman}},\ }\href
  {https://www.wiley.com/en-us/Absorption+and+Scattering+of+Light+by+Small+Particles-p-9780471293408}
  {\emph {\bibinfo {title} {{Absorption and Scattering of Light by Small
  Particles}}}}\ (\bibinfo  {publisher} {Wiley},\ \bibinfo {address} {Hoboken,
  NJ, USA},\ \bibinfo {year} {1998})\BibitemShut {NoStop}%
\bibitem [{sup()}]{supp_mat}%
  \BibitemOpen
  \href@noop {} {\bibinfo {title} {{Supplemental Material}}}\BibitemShut
  {NoStop}%
\bibitem [{\citenamefont {Bernal~Arango}\ \emph {et~al.}(2014)\citenamefont
  {Bernal~Arango}, \citenamefont {Coenen},\ and\ \citenamefont
  {Koenderink}}]{BernalArango2014May}%
  \BibitemOpen
  \bibfield  {author} {\bibinfo {author} {\bibfnamefont {F.}~\bibnamefont
  {Bernal~Arango}}, \bibinfo {author} {\bibfnamefont {T.}~\bibnamefont
  {Coenen}},\ and\ \bibinfo {author} {\bibfnamefont {A.~F.}\ \bibnamefont
  {Koenderink}},\ }\bibfield  {title} {\bibinfo {title} {{Underpinning
  Hybridization Intuition for Complex Nanoantennas by Magnetoelectric
  Quadrupolar Polarizability Retrieval}},\ }\href
  {https://doi.org/10.1021/ph5000133} {\bibfield  {journal} {\bibinfo
  {journal} {ACS Photonics}\ }\textbf {\bibinfo {volume} {1}},\ \bibinfo
  {pages} {444} (\bibinfo {year} {2014})}\BibitemShut {NoStop}%
\bibitem [{\citenamefont {Mun}\ and\ \citenamefont {Rho}(2019)}]{Mun2019May}%
  \BibitemOpen
  \bibfield  {author} {\bibinfo {author} {\bibfnamefont {J.}~\bibnamefont
  {Mun}}\ and\ \bibinfo {author} {\bibfnamefont {J.}~\bibnamefont {Rho}},\
  }\bibfield  {title} {\bibinfo {title} {{Importance of higher-order multipole
  transitions on chiral nearfield interactions}},\ }\href
  {https://doi.org/10.1515/nanoph-2019-0046} {\bibfield  {journal} {\bibinfo
  {journal} {P. Soc. Photo-opt. Ins.}\ }\textbf {\bibinfo {volume} {8}},\
  \bibinfo {pages} {941} (\bibinfo {year} {2019})}\BibitemShut {NoStop}%
\bibitem [{\citenamefont {Das}\ \emph {et~al.}(2015{\natexlab{b}})\citenamefont
  {Das}, \citenamefont {Iyer}, \citenamefont {DeCrescent},\ and\ \citenamefont
  {Schuller}}]{Das2015-Beamengineeringfor}%
  \BibitemOpen
  \bibfield  {author} {\bibinfo {author} {\bibfnamefont {T.}~\bibnamefont
  {Das}}, \bibinfo {author} {\bibfnamefont {P.~P.}\ \bibnamefont {Iyer}},
  \bibinfo {author} {\bibfnamefont {R.~A.}\ \bibnamefont {DeCrescent}},\ and\
  \bibinfo {author} {\bibfnamefont {J.~A.}\ \bibnamefont {Schuller}},\
  }\bibfield  {title} {\bibinfo {title} {{Beam engineering for selective and
  enhanced coupling to multipolar resonances}},\ }\href
  {https://doi.org/10.1103/PhysRevB.92.241110} {\bibfield  {journal} {\bibinfo
  {journal} {Phys. Rev. B}\ }\textbf {\bibinfo {volume} {92}},\ \bibinfo
  {pages} {241110} (\bibinfo {year} {2015}{\natexlab{b}})}\BibitemShut
  {NoStop}%
\bibitem [{\citenamefont {Melik-Gaykazyan}\ \emph {et~al.}(2018)\citenamefont
  {Melik-Gaykazyan}, \citenamefont {Kruk}, \citenamefont {Camacho-Morales},
  \citenamefont {Xu}, \citenamefont {Rahmani}, \citenamefont {Zangeneh~Kamali},
  \citenamefont {Lamprianidis}, \citenamefont {Miroshnichenko}, \citenamefont
  {Fedyanin}, \citenamefont {Neshev},\ and\ \citenamefont
  {Kivshar}}]{Melik-Gaykazyan2018-SelectiveThird-Harmo}%
  \BibitemOpen
  \bibfield  {author} {\bibinfo {author} {\bibfnamefont {E.~V.}\ \bibnamefont
  {Melik-Gaykazyan}}, \bibinfo {author} {\bibfnamefont {S.~S.}\ \bibnamefont
  {Kruk}}, \bibinfo {author} {\bibfnamefont {R.}~\bibnamefont
  {Camacho-Morales}}, \bibinfo {author} {\bibfnamefont {L.}~\bibnamefont {Xu}},
  \bibinfo {author} {\bibfnamefont {M.}~\bibnamefont {Rahmani}}, \bibinfo
  {author} {\bibfnamefont {K.}~\bibnamefont {Zangeneh~Kamali}}, \bibinfo
  {author} {\bibfnamefont {A.}~\bibnamefont {Lamprianidis}}, \bibinfo {author}
  {\bibfnamefont {A.~E.}\ \bibnamefont {Miroshnichenko}}, \bibinfo {author}
  {\bibfnamefont {A.~A.}\ \bibnamefont {Fedyanin}}, \bibinfo {author}
  {\bibfnamefont {D.~N.}\ \bibnamefont {Neshev}},\ and\ \bibinfo {author}
  {\bibfnamefont {Y.~S.}\ \bibnamefont {Kivshar}},\ }\bibfield  {title}
  {\bibinfo {title} {{Selective Third-Harmonic Generation by Structured Light
  in Mie-Resonant Nanoparticles}},\ }\href
  {https://doi.org/10.1021/acsphotonics.7b01277} {\bibfield  {journal}
  {\bibinfo  {journal} {ACS Photonics}\ }\textbf {\bibinfo {volume} {5}},\
  \bibinfo {pages} {728} (\bibinfo {year} {2018})}\BibitemShut {NoStop}%
\bibitem [{\citenamefont {Koshelev}\ \emph {et~al.}(2020)\citenamefont
  {Koshelev}, \citenamefont {Kruk}, \citenamefont {Melik-Gaykazyan},
  \citenamefont {Choi}, \citenamefont {Bogdanov}, \citenamefont {Park},\ and\
  \citenamefont {Kivshar}}]{Koshelev2020Jan}%
  \BibitemOpen
  \bibfield  {author} {\bibinfo {author} {\bibfnamefont {K.}~\bibnamefont
  {Koshelev}}, \bibinfo {author} {\bibfnamefont {S.}~\bibnamefont {Kruk}},
  \bibinfo {author} {\bibfnamefont {E.}~\bibnamefont {Melik-Gaykazyan}},
  \bibinfo {author} {\bibfnamefont {J.-H.}\ \bibnamefont {Choi}}, \bibinfo
  {author} {\bibfnamefont {A.}~\bibnamefont {Bogdanov}}, \bibinfo {author}
  {\bibfnamefont {H.-G.}\ \bibnamefont {Park}},\ and\ \bibinfo {author}
  {\bibfnamefont {Y.}~\bibnamefont {Kivshar}},\ }\bibfield  {title} {\bibinfo
  {title} {{Subwavelength dielectric resonators for nonlinear nanophotonics}},\
  }\href {https://doi.org/10.1126/science.aaz3985} {\bibfield  {journal}
  {\bibinfo  {journal} {Science}\ }\textbf {\bibinfo {volume} {367}},\ \bibinfo
  {pages} {288} (\bibinfo {year} {2020})}\BibitemShut {NoStop}%
\bibitem [{\citenamefont {Palik}(1998)}]{palik1998handbook}%
  \BibitemOpen
  \bibfield  {author} {\bibinfo {author} {\bibfnamefont {E.~D.}\ \bibnamefont
  {Palik}},\ }\href@noop {} {\emph {\bibinfo {title} {Handbook of optical
  constants of solids}}},\ Vol.~\bibinfo {volume} {3}\ (\bibinfo  {publisher}
  {Academic press},\ \bibinfo {year} {1998})\BibitemShut {NoStop}%
\bibitem [{\citenamefont {Varault}\ \emph {et~al.}(2013)\citenamefont
  {Varault}, \citenamefont {Rolly}, \citenamefont {Boudarham}, \citenamefont
  {Dem{\ifmmode\acute{e}\else\'{e}\fi}sy}, \citenamefont {Stout},\ and\
  \citenamefont {Bonod}}]{Varault2013-Multipolareffectson}%
  \BibitemOpen
  \bibfield  {author} {\bibinfo {author} {\bibfnamefont {S.}~\bibnamefont
  {Varault}}, \bibinfo {author} {\bibfnamefont {B.}~\bibnamefont {Rolly}},
  \bibinfo {author} {\bibfnamefont {G.}~\bibnamefont {Boudarham}}, \bibinfo
  {author} {\bibfnamefont {G.}~\bibnamefont
  {Dem{\ifmmode\acute{e}\else\'{e}\fi}sy}}, \bibinfo {author} {\bibfnamefont
  {B.}~\bibnamefont {Stout}},\ and\ \bibinfo {author} {\bibfnamefont
  {N.}~\bibnamefont {Bonod}},\ }\bibfield  {title} {\bibinfo {title}
  {{Multipolar effects on the dipolar polarizability of magneto-electric
  antennas}},\ }\href {https://doi.org/10.1364/OE.21.016444} {\bibfield
  {journal} {\bibinfo  {journal} {Opt. Express}\ }\textbf {\bibinfo {volume}
  {21}},\ \bibinfo {pages} {16444} (\bibinfo {year} {2013})}\BibitemShut
  {NoStop}%
\bibitem [{\citenamefont {Achouri}\ and\ \citenamefont
  {Martin}(2021)}]{Achouri2021Feb}%
  \BibitemOpen
  \bibfield  {author} {\bibinfo {author} {\bibfnamefont {K.}~\bibnamefont
  {Achouri}}\ and\ \bibinfo {author} {\bibfnamefont {O.~J.~F.}\ \bibnamefont
  {Martin}},\ }\bibfield  {title} {\bibinfo {title} {{Extension of Lorentz
  Reciprocity and Poynting Theorems for Spatially Dispersive Media with
  Quadrupolar Responses}},\ }\href {https://arxiv.org/abs/2102.08197v1}
  {\bibfield  {journal} {\bibinfo  {journal} {arXiv:2102.08197}\ } (\bibinfo
  {year} {2021})},\ \Eprint {https://arxiv.org/abs/2102.08197} {2102.08197}
  \BibitemShut {NoStop}%
\bibitem [{\citenamefont {Mun}\ \emph {et~al.}(2020)\citenamefont {Mun},
  \citenamefont {So}, \citenamefont {Jang},\ and\ \citenamefont
  {Rho}}]{Mun2020May}%
  \BibitemOpen
  \bibfield  {author} {\bibinfo {author} {\bibfnamefont {J.}~\bibnamefont
  {Mun}}, \bibinfo {author} {\bibfnamefont {S.}~\bibnamefont {So}}, \bibinfo
  {author} {\bibfnamefont {J.}~\bibnamefont {Jang}},\ and\ \bibinfo {author}
  {\bibfnamefont {J.}~\bibnamefont {Rho}},\ }\bibfield  {title} {\bibinfo
  {title} {{Describing Meta-Atoms Using the Exact Higher-Order Polarizability
  Tensors}},\ }\href {https://doi.org/10.1021/acsphotonics.9b01776} {\bibfield
  {journal} {\bibinfo  {journal} {ACS Photonics}\ }\textbf {\bibinfo {volume}
  {7}},\ \bibinfo {pages} {1153} (\bibinfo {year} {2020})}\BibitemShut
  {NoStop}%
\bibitem [{\citenamefont {Doost}\ \emph {et~al.}(2014)\citenamefont {Doost},
  \citenamefont {Langbein},\ and\ \citenamefont {Muljarov}}]{Doost2014Jul}%
  \BibitemOpen
  \bibfield  {author} {\bibinfo {author} {\bibfnamefont {M.~B.}\ \bibnamefont
  {Doost}}, \bibinfo {author} {\bibfnamefont {W.}~\bibnamefont {Langbein}},\
  and\ \bibinfo {author} {\bibfnamefont {E.~A.}\ \bibnamefont {Muljarov}},\
  }\bibfield  {title} {\bibinfo {title} {{Resonant-state expansion applied to
  three-dimensional open optical systems}},\ }\href
  {https://doi.org/10.1103/PhysRevA.90.013834} {\bibfield  {journal} {\bibinfo
  {journal} {Phys. Rev. A}\ }\textbf {\bibinfo {volume} {90}},\ \bibinfo
  {pages} {013834} (\bibinfo {year} {2014})}\BibitemShut {NoStop}%
\end{thebibliography}%

\end{document}